\documentclass[fleqn,usenatbib]{mnras}

\usepackage{newtxtext,newtxmath}

\usepackage[T1]{fontenc}

\DeclareRobustCommand{\VAN}[3]{#2}
\let\VANthebibliography\thebibliography
\def\thebibliography{\DeclareRobustCommand{\VAN}[3]{##3}\VANthebibliography}

\usepackage{graphicx}	
\usepackage{amsmath}	
\usepackage{subcaption}
\usepackage{caption}
\usepackage[T1]{fontenc}

\usepackage{placeins}
\usepackage{afterpage}

\newcommand{\equarefalt}[1]{Equation~\ref{#1}}
\newcommand{\figref}[1]{Fig.~\ref{#1}}
\newcommand{\tabref}[1]{Table~\ref{#1}}

\title[Bar-induced substructures in the stellar halo]{Stirring Things Up: Bar-induced substructures in the stellar halo of a cosmological Milky Way analogue}

\author[T. Tomlinson et al.]{Thomas Tomlinson$^{1}$\thanks{E-mail: thomas.j.tomlinson@durham.ac.uk},
Francesca Fragkoudi$^{1}$,
Andreia Carrillo$^{1, 2, 3}$,
Azadeh Fattahi$^{1,4}$,
\newauthor
Paula Gherghinescu$^{1}$,
Alis Deason$^{1,2}$, 
Rüdiger Pakmor$^{5}$, 
Robert J. J. Grand$^{6}$, 
\newauthor
Facundo A. Gómez$^{7}$, 
Freeke van de Voort$^{8}$, 
Rebekka Bieri$^{9}$\\
$^{1}$Institute for Computational Cosmology, Department of Physics, University of Durham, South Road, Durham DH1 3LE, UK\\
$^{2}$Centre for Extragalactic Astronomy, Department of Physics, University of Durham, South Road, Durham DH1 3LE, UK\\
$^{3}$Department of Physics and Astronomy, Carleton College, 1 North College St., Northfield, MN 55057, USA\\
$^{4}$The Oskar Klein Centre, Department of Physics, Stockholm University, Albanova University Center, 106 91 Stockholm, Sweden\\
$^{5}$Max-Planck-Institut für Astrophysik, Karl-Schwarzschild-Str. 1, D-85748, Garching, Germany\\
$^{6}$Astrophysics Research Institute, Liverpool John Moores University, 146 Brownlow Hill, Liverpool, L3 5RF, UK\\
$^{7}$Departamento de Astronomía, Universidad de La Serena, Av. Juan Cisternas 1200 Norte, La Serena, Chile\\
$^{8}$Cardiff Hub for Astrophysics Research and Technology, School of Physics and Astronomy, Cardiff University, Queen’s Buildings, Cardiff CF24 3AA, UK\\
$^{9}$Institut für Astrophysik, Universität Zürich, Winterthurerstrasse 190, 8057 Zürich, Switzerland
}
\date{Accepted XXX. Received YYY; in original form ZZZ}

\pubyear{\the\year{}}

\begin{document}
\label{firstpage}
\maketitle

\begin{abstract}
The stellar halo of the Milky Way contains the remnants of past accretion events, which could be detectable as substructures in the classical integrals of motion space, such as energy and angular momentum ($E-L_z$). However, our galaxy also contains a non-axisymmetric stellar bar, which traps stars in resonant orbits, leading to substructures in phase-space. Using a high-resolution magneto-hydrodynamic cosmological zoom-in simulation of a Milky Way analogue, we explore the connection between the bar and the accreted stellar halo. We find that the bar induces prominent substructures, or `ridges’, in $E-L_z$, caused by the resonances. The most pronounced of these is caused by the corotation and the retrograde 1:1 resonances, with weaker ridges visible due to the prograde 1:1 and outer Lindblad resonance. The ridges are present across much of the stellar halo, with variations in radius due to the morphology of different orbital families. We explore the scattering of orbits at the resonances, finding that stars trapped at the 1:1 retrograde resonance become more circularised and have more negative angular momentum. Additionally, stars can move between the corotation and retrograde 1:1 families, thus alternating between prograde and retrograde motion. Due to these scatterings and the pre-existing metallicity gradients in the accreted population, the bar-induced substructures have distinct metallicities compared to stars in the surrounding phase-space. Our results suggest the need for caution when searching the Milky Way stellar halo for accreted substructures in both integral of motions and chemical spaces, since these can be induced by internal perturbations.
\end{abstract}




\section{Introduction}
\label{sec:intro}
The notion that the Milky Way is a barred spiral galaxy was first proposed by \citet{vaucouleurs_1964} and further supported by \citet{binney_1991}, \citet{blitz_1991} and \citet{englmaier_1999}, among others. The bar is a feature the Milky Way has in common with around two thirds of spiral galaxies in the local universe \citep{menendez_2007, sheth_2008}, with similar fractions already found in the 1960s by \citet{Vaucouleurs_1963}. \par

Bars are known to drive the exchange of angular momentum within galaxies \citep{lynden_bell_1972} and to thus couple the evolution of stellar discs to their surrounding haloes. 
Early experiments with numerical simulations over half a century ago by \citet{ostriker_1973} showed that a stellar disc embedded in the potential of a massive halo can be stabilised against the formation of a bar. Subsequent studies showed that a `live' dark matter halo (i.e., one in which dark matter is modelled as $N$-body particles that self-consistently respond to changes in the potential) can absorb angular momentum emitted by the disc through the bar \citep{athanassoula_2002}. This transfer occurs most efficiently at the resonances. Orbits are said to be in resonance with the bar when they satisfy the condition, \par

\begin{equation}
    m(\Omega_{\phi} - \Omega_{\mathrm{bar}}) + l \; \Omega_R = 0,
    \label{eq:resonance}
\end{equation}

\noindent where $\Omega_\phi$ and $\Omega_R$ are the orbital azimuthal and radial frequency respectively, $\Omega_{\rm bar}$ is the pattern speed of the bar, and $l$ and $m$ are integers. Resonances are identified via combinations of $l$ and $m$, with some of the most notable ones being the inner Lindblad resonance (ILR; $l$/$m$ = -1/2), the corotation resonance (CR; $l$=0), the outer Lindblad resonance (OLR; $l$/$m$ = 1/2), and the 1:1 resonance ($l$/$m$ = 1). 
Stars trapped at the ILR are within the bar; indeed it has been shown that the so-called $x_1$ bar-supporting family of orbits are limited by the corotation radius in extent \citep{Contopoulos_1989}. Stars trapped at the CR and OLR are present outside the bar, while the 1:1 orbits have both a prograde and retrograde component, and are found outside and inside the corotation radius, respectively \citep{Contopoulos_1989}.
Angular momentum is emitted by stars trapped at the ILR, while the CR and OLR absorb angular momentum \citep{lynden_bell_1972}. The amount of angular momentum that can be exchanged at the resonances relies on a delicate balance between this `absorbing' and `emitting' resonant material \citep{athanassoula_2003}. 

While the aforementioned theoretical studies have explored the interaction between the stellar disc and the dark matter halo via the bar, little attention has been paid to the consequences of this angular momentum transfer on the stellar halo.
This changed with the recent work by \citet{dillamore_2023, dillamore_2024}, who used test particle simulations to show that stars in the stellar halo, which are trapped in resonance by the bar, will form overdensities in phase space and in the space of integrals of motion, such as energy, $E$, and the vertical component of the angular momentum, $L_z$. Furthermore, by using data from Gaia DR3 \citep{gaia_2023}, they find an overdensity in $E-L_z$ in the stellar halo of the Milky Way, which they deduce is caused by stars trapped in corotation with the bar. 

Previously, substructures in $E-L_z$ space in the stellar halo have been associated with accretion of satellite galaxies, as established by the work of \citet{helmi_2000}. These authors used numerical simulations to show that in the case of a static axisymmetric potential, overdensities in $E-L_z$ space are largely conserved, even after several Gyrs and after the satellite has been completely disrupted in configuration space.
The last decade has seen many advancements in the study of the Milky Way's stellar halo and the search for substructures in the Galaxy, which have largely been made possible by the launch of the Gaia satellite and its more than 1.8 billion measurements of positions on the sky, parallaxes, and proper motions of stars in the Milky Way \citep{gaia_2016, gaia_2023}. In synergy with dedicated large spectroscopic surveys such as APOGEE \citep{majewski_2017}, DESI \citep{cooper_2023}, and upcoming surveys such as 4MOST \citep{de_jong_2019} among others, it is now possible to study the dynamics and chemistry of the stellar halo of the Milky Way on a star-by-star basis for a significant fraction of stars. Among the first discoveries enabled by Gaia was the discovery of a massive merger which dominates the halo of the Milky Way within 15-20 kpc from the centre, dubbed the Gaia Sausage/Enceladus (GSE) merger \citep{belokurov_2018, helmi_2018, haywood_2018, naidu_2020}. In $E-L_z$ space, the  GSE occupies a large region in the area around $L_z = 0$. Several other overdensities in $E-L_z$ space have been associated with other merger events. Some examples of these postulated mergers are Thamnos, a retrograde structure \citep{koppelman_2019}, Kraken, a proposed early accretion event found at low energy \citep{kruijssen_2020} and Heracles \citep{horta_2021, horta_2024}.

However, while $E-L_z$ are approximately conserved quantities, several studies have pointed out that caution needs to be exercised when equating substructures in this space to accretion events. For example, \citet{jean_baptiste_2017} showed that, in simulations with a live potential, even a single accretion event can cause multiple overdensities in $E-L_z$ space. As the merging progenitor is redistributed in $E-L_z$ space, sinking to lower energies due to dynamical friction, separate clumps form as it loses stars at consecutive pericentric passages. Furthermore, in-situ stars can also form overdensities by being redistributed after being heated by a merger.

In addition to this, in non-axisymmetric potentials, such as a galaxy with a bar, like the Milky Way, the energy and angular momentum are no longer conserved along a given orbit. However, the combination of the two via the Jacobi Energy, $E_J$,

 \begin{equation}
     E_J = E - \Omega_{\mathrm{bar}}L_z,
    \label{eq:Ejacobi}
\end{equation}

\noindent is conserved \citep{binney_2008}.
Thus, substructures caused by mergers can be blurred by the bar, or indeed new substructures can be induced, in the classical integrals of motion space $E-L_z$. This adds complications in identifying past mergers of the Milky Way. 

While previous works have used test particle simulations to explore bar-induced resonances \citep{dillamore_2023,dillamore_2024}, these do not provide information on the assembly of the galaxy. Cosmological simulations model the formation and evolution of galaxies self-consistently within the $\mathrm{\Lambda CDM}$ context; we can separate stars into those that are born in the main galaxy (in-situ) as well as those that have been accreted over time (ex-situ), while also incorporating information about the ages and elemental abundances of stellar populations in the halo.
We leverage these aspects in our current study, using a newly-developed high-resolution (800 $\mathrm{M_{\sun}}$) cosmological zoom-in simulation of a Milky Way analogue with a stellar bar. The bar in this simulation, as we shall see, causes prominent overdensities in the stellar halo of the system. We investigate which resonances cause these overdensities in $E-L_z$ space and what they might tell us about the evolution of our Galaxy and the search for substructures, both secular and merger-induced. 

The paper is structured as follows: \autoref{sec:Simulation and Methods} describes the simulation and analysis tools we use; \autoref{sec:Results} shows the results of our analysis, showing the origin of the substructures in $E-L_z$, how these relate to orbits in configuration space and their properties in terms of chemical abundances; \autoref{sec:Discussion} discusses the importance of the 1:1 retrograde resonance and how orbits are scattered by this resonance; \autoref{sec:Conclusions} then concludes and summarises our results.

\section{Simulation and Methods} \label{sec:Simulation and Methods}

\begin{figure}
	\includegraphics[width=\columnwidth]{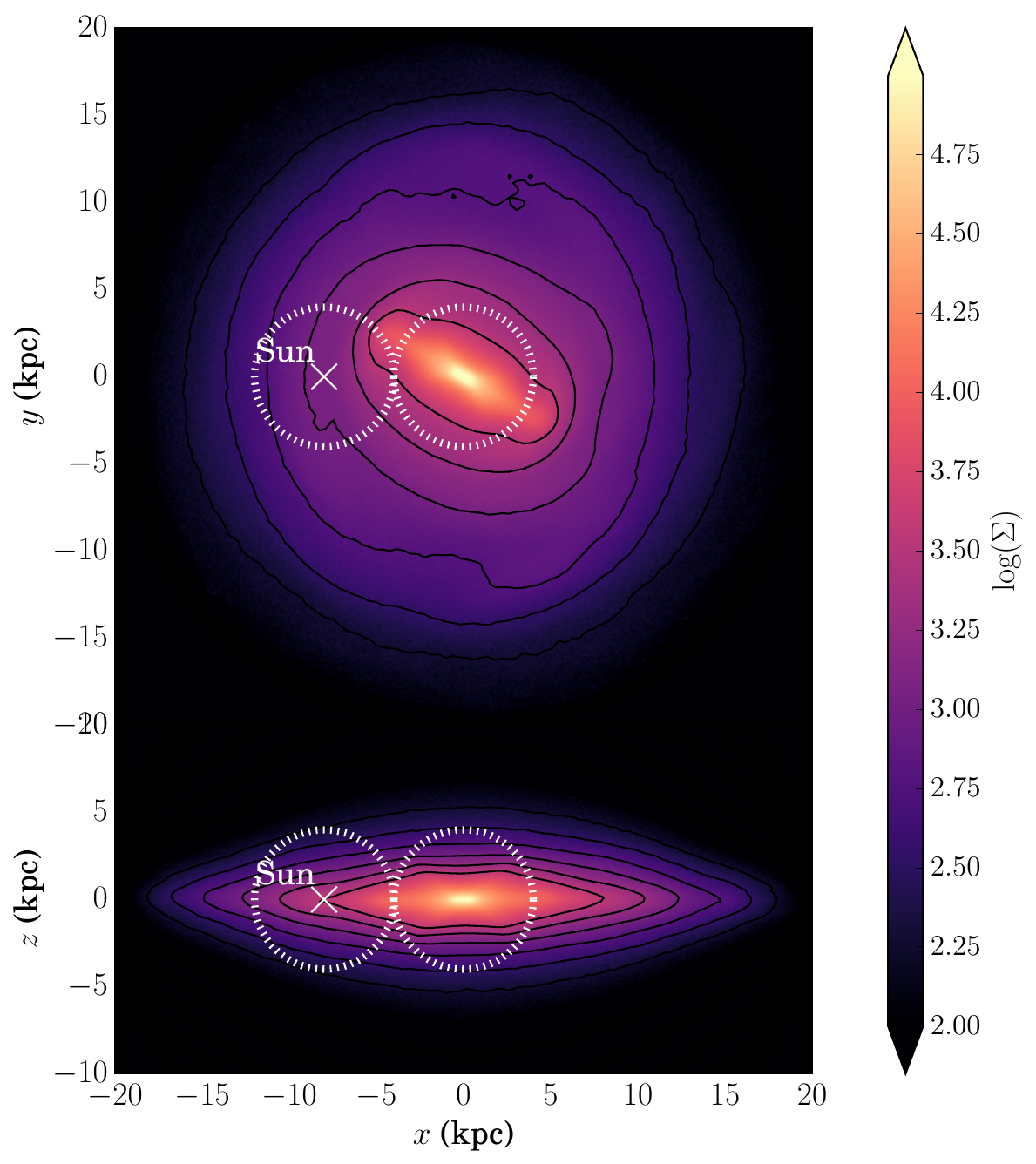}
    \caption{Logarithmic stellar surface density projection in x-y and x-z of the simulated halo, Auriga 18. We define a solar neighbourhood in this simulation following observations, where the sun is located at x = -8 kpc, 30 degrees behind the bar. The circles around the centre of the galaxy and the sun indicate the two regions studied, each with a radius of 4 kpc.}
    \label{fig:sdens}
\end{figure}

\subsection{Auriga Superstars Milky Way analogue}
The simulated galaxy we explore in this work is part of the newly developed Auriga Superstars suite of simulations (\citealt{grand_2023}; \citealt{pakmor_2025}; Fragkoudi et al., in prep.). These are cosmological zoom-in simulations based on the original Auriga suite of simulations \citet{grand_2017, grand_2024} tailored for studies of stellar dynamics. They are re-simulated with 64 times better resolution in terms of the stellar mass ($800 \;M_{\sun}$ resolution), and 8 times better resolution for the dark matter particles ($5\times10^4 \;M_{\sun}$ resolution), as compared to the original standard resolution Auriga simulations. A full description of the Superstars method is presented in \citet{pakmor_2025} and the suite of simulations will be presented in Fragkoudi et al. (in prep.). Here we give a brief summary of the basic properties of the simulations.

The original Auriga simulations are sampled from a cubic volume with side lengths of 100\;Mpc taken from the EAGLE project \citep{schaye_2015}. The simulations are run with the \textsc{arepo} code \citep{springel_2010, pakmor_2016}, in a standard $\mathrm{\Lambda CDM}$ cosmology, with parameters $\Omega_{\rm m} = 0.307$, $\Omega_{\rm{b}} = 0.048$, $\Omega_{\Lambda} = 0.693$ and $H_{0} \mathrm{= 0.6777 \; km \; s^{-1} \; Mpc^{-1}}$ \citep{planck_2013}. They make use of the zoom-in technique to achieve a baryonic mass resolution of $\sim 5 \times 10^{4} \; M_{\sun}$, with a typical dark matter resolution of $\sim 3 \times 10^{5} \; M_{\sun}$. In this work, we refer to the original Auriga simulations as the "fiducial" or "Level 4" Auriga simulations. The details of the Auriga suite of simulations are described in detail in \citet{grand_2017}. \par 

The Superstars method creates a larger number of lower-mass ($\sim 800 \; M_{\sun}$) star particles from a single gas cell. These stars have the same velocity and position as their parent gas cell with an added random isotropic component based on a Gaussian distribution with a width set by the minimum of the local sound speed and the velocity dispersion of the neighbouring gas cells\footnote{We note that the simulation studied here uses a variation of the Superstars method, with a factor of $\sim 2$ larger kick to the birth velocity to stars. However, we highlight that this does not have a significant impact on the properties of the galaxy -- see the discussion and Appendix A in \citealt{pakmor_2025} for more details.} \citep{pakmor_2025}. Care is taken that these additional random components do not affect the conservation of total momentum. The chemical evolution of the stars is treated in the same way as in the fiducial Auriga simulation. Re-simulating a halo using the Superstars method keeps galaxy properties largely preserved, with variations falling within the range of intrinsic variations found among realisations run on a different machine or with a different random number seed \citep{pakmor_2025_a}. The changes that do occur are smaller than the systematic shifts caused by changing the mass resolution of the gas. For more details on the method, we refer the reader to \citet{pakmor_2025}. \par 

\subsubsection{Halo 18}
In this study we use halo 18 from the fiducial Auriga suite of simulations, re-simulated using the Superstars method. The simulation produces full outputs every 50\,Myrs. Additionally, every 5\,Myrs a `snipshot' with the positions of all particles is created, which we use for our orbital analysis, as well as our estimates of the bar pattern speed. Metal abundances are extracted from the simulation, and we use [Fe/H] values as a proxy for metallicity; we reduce [Fe/H] by 0.5 dex to bring the  metallicities in line with the Milky Way in the Milky Way bulge. The dark matter halo mass is $\sim 1.3\times10^{12}M_{\sun}$ with a stellar mass of $\sim 7\times10^{10}M_{\sun}$. It is a useful analogue of the Milky Way in a number of ways: it is a barred galaxy whose inner regions have chemodynamical properties similar to the Milky Way \citep{fragkoudi_2020}, and it has an analogue for the GSE merger, with a mass and accretion time that fall within the estimates for GSE, and shared kinematic properties \citep{fattahi_2019}.

The bar forms at a lookback time of  $\sim$8\,Gyrs (see also \citealt{merrow_2024}), and has a length of $\sim$5.6 kpc, based on the radius at which the $m=2$ Fourier mode of stars in the disc reaches 60\% of its maximum (see e.g. \citealt{fragkoudi_2021} for the method). This is close to recent estimates of the length of the Milky Way's bar of $\approx5$\,kpc, which varies depending on the method used \citep{portail_2017, hilmi_2020, lucey_2023}. The pattern speed of the bar is $23.0\; \rm km \, s^{-1} \, kpc^{-1}$ at the $z=0$ snapshot, which we calculated from the difference in the phase of the bar in the two most recent simulation snipshots. The corotation radius is $\sim 10$\,kpc, which is slightly larger than estimates for the Milky Way, which place its corotation radius between $4.5-9.6\,$kpc, albeit with significant uncertainty \citep{bland_hawthorn_2016,sanders_2019,chiba_2021,horta_2025}. The bar pattern speed, and therefore the locations of the different resonances, such as the corotation resonance, affect the regions that will be occupied by different families of orbits -- an important caveat to keep in mind when interpreting our findings and how they might correspond to the Milky Way. In Appendix \ref{sec:appendix c} we show an alternative solar neighbourhood at 11.5 kpc. This radius is chosen such that the ratio of the corotation radius over the solar radius is around $8/7$, which is the ratio of these two radii in the Milky Way if we assume that the corotation radius in our Galaxy is around 7\,kpc and the solar radius is 8\,kpc. 

In what follows, we explore two regions of the simulated galaxy, as shown in \figref{fig:sdens}: a solar neighbourhood-like region, i.e. a sphere of radius 4\,kpc at a cylindrical radius of $R$ = 8\,kpc, in the plane of the galaxy, and a sphere of 4\,kpc at the centre of the galaxy. We rotate the bar to have an angle of 30 degrees ahead of the `Sun'-galactic centre line, similar to the Milky Way. The scale radius of the disc of our simulated galaxy is 3.7 kpc, which is on the high end of estimates for the Milky Way scale length (for which studies find values between $2.6-3.9\,$kpc; see e.g. \citealt{bland_hawthorn_2016}). 

In what follows, stars are selected as in-situ or accreted, following the methodology used in \citet{fattahi_2019}. This makes use of the friends-of-friends and \textsc{subfind} algorithms \citep{davis_1985, springel_2001a, springel_2001b}. Star particles are assigned to a progenitor based on what subhalo they are bound to in the snapshot after their formation. Thus, stars formed from gas stripped from a progenitor will be flagged as in-situ if the star that forms is bound to the main Milky Way-like progenitor. Additionally, stars which were accreted before a redshift of $z=3.8$ are also counted as in-situ, because at redshifts earlier than this, identifying the main halo is non-trivial as several progenitors have similar masses. 
In this study we explore the four most massive accretion events in the galaxy, which combined make up more than 90\% of the accreted stellar mass within 20\,kpc of the centre of the halo at $z=0$. In \tabref{tab:1} we list their peak stellar mass, peak halo mass, the fraction of stellar mass from each progenitor to the total accreted population within 20\,kpc, their mean metallicity, infall time and redshift. 

\begin{table}
\centering
\begin{tabular}{ |p{.7cm}|p{1.cm}|p{1.2cm}|p{.7cm}|p{.7cm}|p{.7cm}|p{.7cm}|}
\hline
Merger & Peak stellar mass ($M_{\sun}$) & Peak halo mass ($M_{\sun}$) & Accreted \nobreak{fraction} & Mean [Fe/H] & Infall time (Gyrs) & Infall redshift\\
\hline
M1 & $1.5\times10^{8}$ & $3.0\times10^{9}$ & 0.1 & -1.16 & 12.52 & 4.6\\
M2 & $1.9\times10^{8}$ & $9.6\times10^{9}$ & 0.1 & -1.18 & 12.11 & 3.7\\
M3 & $5.6\times10^{8}$ & $1.6\times10^{10}$ & 0.12 & -0.89 & 9.47 & 1.5\\
M4 & $1.5\times10^{9}$ & $3.6\times10^{10}$ & 0.61 & -0.63 & 9.29 & 1.4\\
\hline
\end{tabular}
\caption{\label{tab:1} Properties of the four main mergers in simulation. The peak stellar and halo mass is given by \textsc{subfind}. The accreted fraction describes how many accreted stars within 20\,kpc of the centre of the main halo come from each merger. The mean metallicity given is for stars found within 20\,kpc of centre of the main halo at $z=0$, calculated as the mean of values of [Fe/H] after taking the logarithm. The infall time (and redshift) refers to the first time the satellite crossed $\mathrm{R_{200}}$ of the main halo. The time is given as the lookback time. The start of the simulation occurs at a lookback time of 13.83\,Gyrs.}
\end{table}

\subsection{Calculating orbital frequencies}
\label{sec:orbfreq}

We calculate the orbital frequencies of stars in the simulation directly from the snipshot outputs of the simulation, which are saved every 5\,Myr. This is done for stars within $r=20\,\mathrm{kpc}$ for a time interval of 2\,Gyr before $z=0$. The time period is chosen to make sure that several stellar orbits are captured for the majority of stars in this region. The frequencies are calculated using the \texttt{scipy.fft} fast Fourier transform (FFT) package from \texttt{scipy} \citep{virtanen_2020}. An FFT is performed on the cylindrical coordinates $R$, $\phi$, and $z$ of the star in an inertial frame of reference in order to obtain the orbital frequencies $\Omega_R$, $\Omega_{\phi}$, and $\Omega_z$. The Fourier transform is performed using a window function of the form:

\begin{equation}
    \chi (t) = 1 + \mathrm{cos}(\frac{2 \pi t}{T}),
    \label{eq:window}
\end{equation}

\noindent where $t$ is the time of each snipshot and $T$ is the time period over which the Fourier transform is performed, to account for the fact that the FFT is carried out over a finite time interval (see \citealt{beraldo_2023} for a more detailed discussion on the use of such a window function). Our frequency analysis is limited by the number of snipshots and time period over which the FFT is performed; at the high end this is given by the Nyquist frequency of 628 rad $\mathrm{Gyr^{-1}}$ and at the low end by a minimum frequency of 3.1 rad $\mathrm{Gyr^{-1}}$.  After performing the FFT, we then obtain the frequency associated with the largest peak as a proxy for the fundamental frequency of the orbit \citep{binney_1982}. We consider peaks of frequencies larger than 3.1 rad $\mathrm{Gyr^{-1}}$. 
We use a convention in which stars whose mean angular momentum over the last 2\,Gyrs is predominantly negative, have a negative $\Omega_\phi$.
To identify orbital resonances, we use the variable $r_\Omega$,

\begin{equation}
    r_{\Omega} = \frac{\Omega_{\phi} - \Omega_{\mathrm{bar}}}{\Omega_R}.
    \label{eq:resonance_r}
\end{equation}

\noindent In this notation the ILR ($l/m$ = -1/2) corresponds to $r_\Omega$ = 0.5, the corotation resonance ($l$=0) to $r_\Omega = 0$, the OLR ($l/m$ = 1/2) to $r_\Omega = -0.5$ and the 1:1 resonance ($l/m$ = 1) to $r_\Omega = -1$. In what follows, we refer to particular resonant families by either their $l/m$ notation or their $r_{\Omega}$ value. \par

The pattern speed of the bar slows down slightly in the time period of our frequency analysis. In the 2\,Gyr interval of our frequency analysis it changes from 26.8 $\mathrm{rad\;Gyr^{-1}}$ to 23.5 $\mathrm{rad\;Gyr^{-1}}$ at the end of the simulation. For the purposes of determining the orbital resonances, we take a bar pattern speed of 25.1 $\mathrm{rad\;Gyr^{-1}}$, which is obtained by performing an FFT on the phase of the bar.

\subsection{Integrals of motion space}

In what follows, we explore how stars cluster together in both the classical integrals of motion space, $E-L_z$, and in action space, i.e. $J_\phi$, $J_R$, $J_z$.
The specific energy, ${E} \; = \; v^{2}/2 + \Phi(x, y, z)$,  is calculated for each particle in the simulation, where $v$ is the modulus of the velocity in a frame centred on the galaxy's centre of mass, subtracting the centre of mass motion of the halo. The potential, $\Phi$, of each particle is extracted from the simulation using a tree code algorithm as described in \citet{springel_2010}. We additionally shift the potential, such that the potential of particles at $\mathrm{R_{200}}$ (i.e., at the virial radius), is equal to zero. The virial radius is defined to be the radius of a sphere in which the mean matter density is 200 times the critical density, $\rho_{\mathrm{crit}} = 3H^2(z)/(8\pi G)$. The specific angular momentum is calculated by $\hat{\textbf{k}} \; \mathrm{L_z}\; = \textbf{r}_{xy} \; \times \; \textbf{v}_{xy}$. 

In a time-independent and axisymmetric potential, the axisymmetric actions, $J_{R}$, $J_{\phi}$, $J_z$ are integrals of motion, meaning they are conserved along a star's orbit \citep{binney_2008}. Each action is associated with a fundamental frequency, $\mathrm{\Omega_R, \; \Omega_{\phi}, \; and \; \Omega_{z}}$. The radial action $J_R$ measures how eccentric an orbit is, describing how much a star oscillates around a guiding radius of a perfectly circular orbit with the same angular momentum. The vertical action $J_z$ measures how far a star deviates from the plane of the galaxy, and the azimuthal action $J_\phi$ -- which is equal to $L_z$ by definition in an axisymmetric system -- is a measure of the prograde and retrograde motion in the plane\footnote{Strictly speaking, the formalism of computing axisymmetric actions, using the St{\"a}ckel fudge approach \citep{binney_2012}, fails at resonances in a non-axisymmetric system, such as that of a barred galaxy. This is because the method relies on the existence of three integrals of motion that are smooth, single-valued, and well-defined. At resonances, those motions become coupled breaking the separability assumption. Because of this, actions are not uniquely defined nor preserved along an orbit.}. We use these axisymmetric actions as approximate orbital labels, calculated for an axisymmetric approximation of the potential in the simulation, as in previous studies (see e.g., \citealt{trick_2019,trick_2021, gomez_2010a, gomez_2010b}).

We calculate this axisymmetric potential approximation directly from the simulation snapshot using the \textsc{agama} package \citep{vasiliev_2019}. The total potential is the sum of the contributions from the dark matter, stars (separated into halo and disc components), and gas. The potentials of the spherical components (dark matter and stellar haloes) are modelled as the sum of different multipole moments, which are expressed as the product of spherical harmonics and an arbitrary function of radius, $\Phi(r, \theta, \phi)=\sum_{j,k} \Phi_{j,k}(r) Y_{j}^{k}(\theta, \phi)$. The radial dependence of these terms is described by an interpolation using a quintic spline, made up of piecewise polynomials, defined by a series of grid nodes evenly spread in $\mathrm{log} \; r$. The flat components (gas and in-situ stars) are modelled via an azimuthal harmonic expansion, using a sum of Fourier terms in the azimuthal angle, i.e., sin($m_i \phi$), cos($m_i \phi$). The coefficients of each term are interpolated on a 2D quintic spline in the ($R, z$) plane. The actions themselves are then calculated using an action finder in \textsc{agama}, which makes use of the St{\"a}ckel fudge method to convert between position/velocity and action/angle variables. For further details on how this is implemented, we direct the reader to the \textsc{agama} documentation \citep{vasiliev_2018}.


\begin{figure*}
    \begin{subfigure}{\textwidth}
        \includegraphics[width=\textwidth]{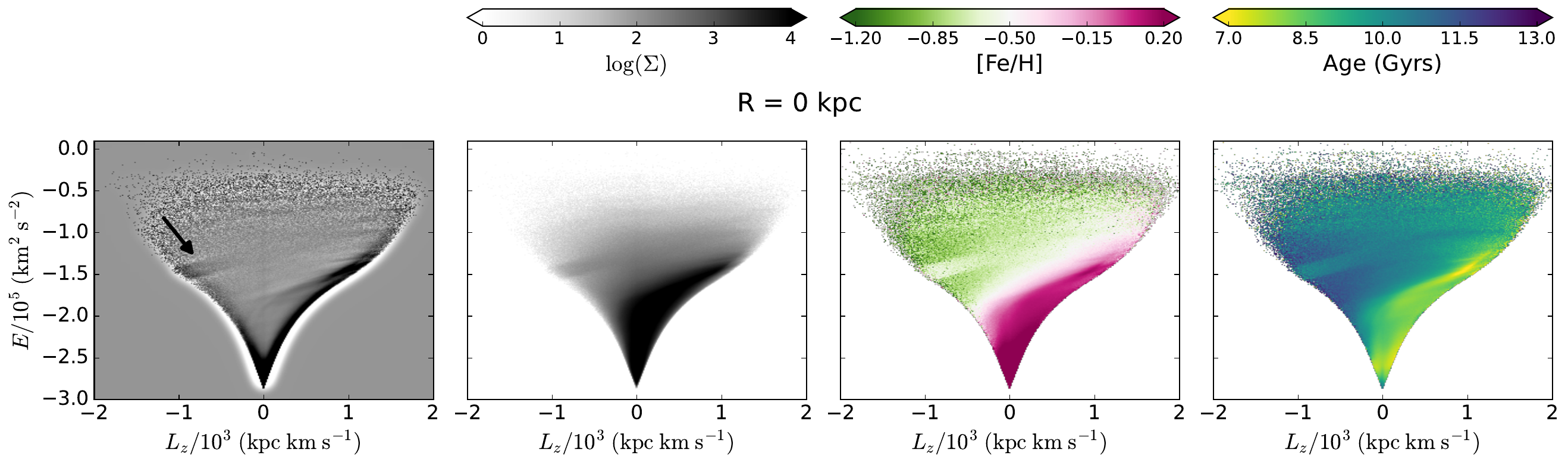}
    \end{subfigure}
    \begin{subfigure}{\textwidth}
        \includegraphics[width=\textwidth]{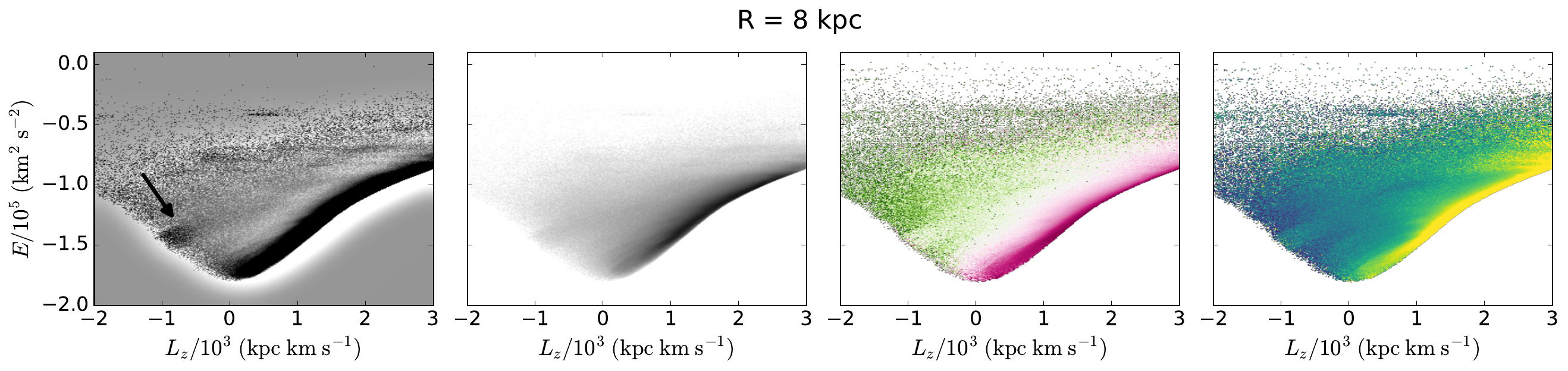}
    \end{subfigure}
    \caption{This figure shows all stars in the simulation in the $E-L_z$ plane in two regions of the simulation, which are used throughout the paper, a 4\,kpc sphere around the centre of the halo (top) and a 4\,kpc sphere around a "solar neighbourhood" (bottom) as illustrated in \figref{fig:sdens}. The leftmost panels show the density of stars in $E-L_z$ space with an unsharp mask applied, the panels in the second column show a logarithmic density projection of the same selection of stars. In the panels in the third column, the bins in $E-L_z$ space are coloured by the mean metallicity of stars inside the bin, and in the rightmost panels the bins are coloured by the mean ages of the stars. In all panels a ridge feature stands out starting at $\mathrm{E \approx -1.5 \times 10^5 \; km^2 \; s^{-2}}$ and $\mathrm{L_z \approx -1 \times 10^3 \; kpc \; km \; s^{-1}}$ (also indicated by arrows in the leftmost panels). Stars inside the ridge tend to have lower ages and higher metallicities than surrounding stars in the same region in $E-L_z$ space.}
    \label{fig:all summary}
\end{figure*}

\begin{figure*}
    \begin{subfigure}{.9\textwidth}
        \includegraphics[width=\textwidth]{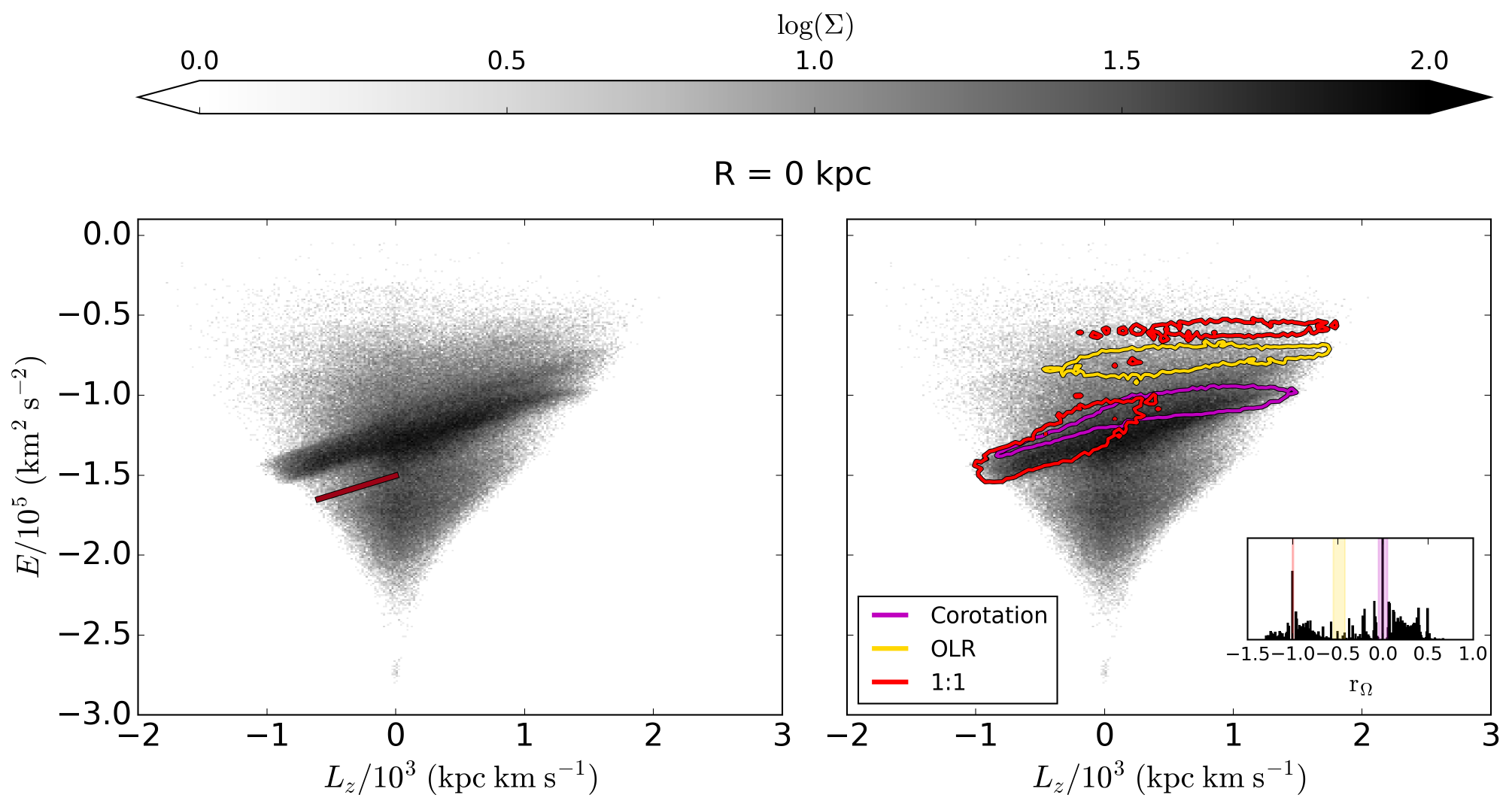}
    \end{subfigure}
    \begin{subfigure}{.9\textwidth}
        \vspace{0em}
        \includegraphics[width=\textwidth]{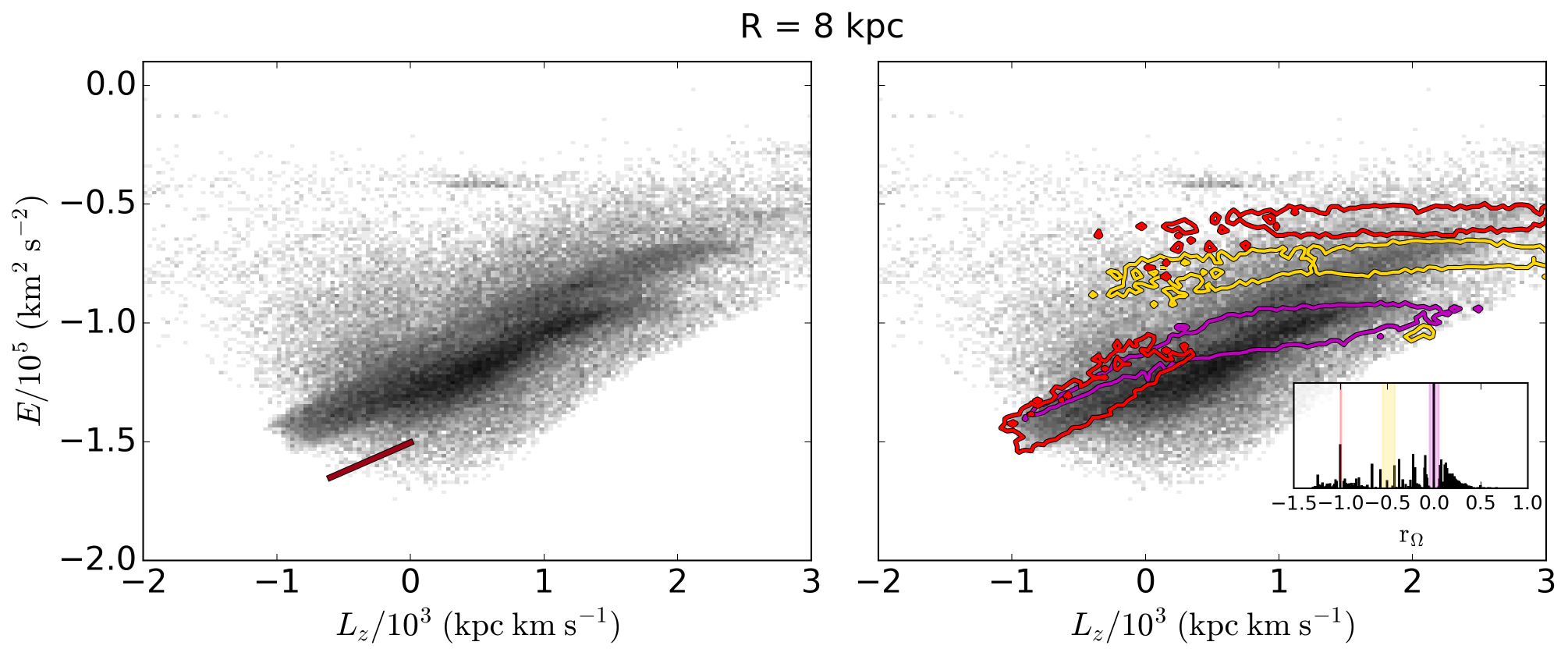}
    \end{subfigure}
    \caption{The logarithmic density distribution of accreted stars in $E-L_z$ space is shown for a central region (top) and a "solar neighbourhood" (bottom), as shown in \figref{fig:sdens}. The plots on the left show these distributions without contours to clearly show the presence of a ridge overdensity, in both spatial regions. Additionally, a line of constant Jacobi energy is plotted in dark red. On the right the same plots are shown with added contours (90th percentile) which show the distribution of stars in three different resonances (corotation - purple, OLR - yellow, 1:1 - red). These colours are used to denote the different resonances throughout the paper. The histogram insets in the two right hand panels show the distribution of stars according to their $\mathrm{r_{\Omega}}$ resonance value; the colours indicate our selection of the resonances, identical to the colours of the contours. We can see that the ridge overdensity is traced mainly by stars in corotation and 1:1 resonance.}
    \label{fig:e-lz_resonanece}
\end{figure*}

\begin{figure*}
    \centering
    \begin{subfigure}{\textwidth}
        \centering
        \includegraphics[width=.9\textwidth]{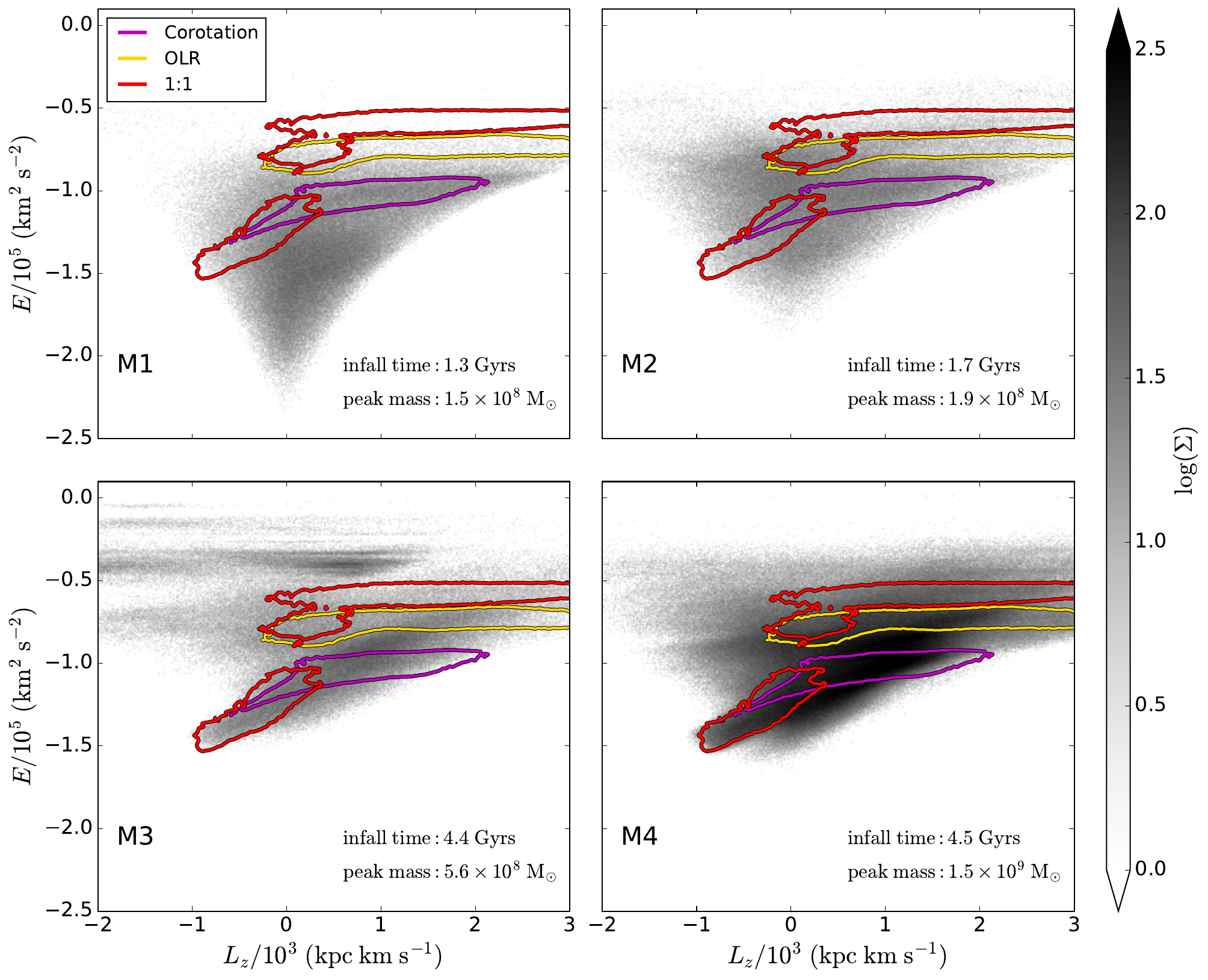}
    \end{subfigure}
    \caption{The four panels show stars from the four main accreted progenitors in the simulation coloured by their logarithmic density in $E-L_z$ space. Shown are all stars from these progenitors within a radius of 20 kpc, not just the central or "solar neighbourhood" regions shown previously. Differently coloured contours (90th percentile) show the distribution of resonant stars for all accreted stars (CR: purple, OLR: yellow, 1:1: red). The mergers are in order of their infall times (left to right, top to bottom). Mergers M1 and M2 merged several Gyrs before bar formation, mergers M3 and M4 merged around the same time shortly before the bar is formed. M4 dominates in terms of its contribution in the region where the ridge is found. However, it is important to note that all four mergers are present inside the ridge.}
    \label{fig:merger_density}
\end{figure*}

\begin{figure*}
    \centering
    \includegraphics[width=\linewidth]{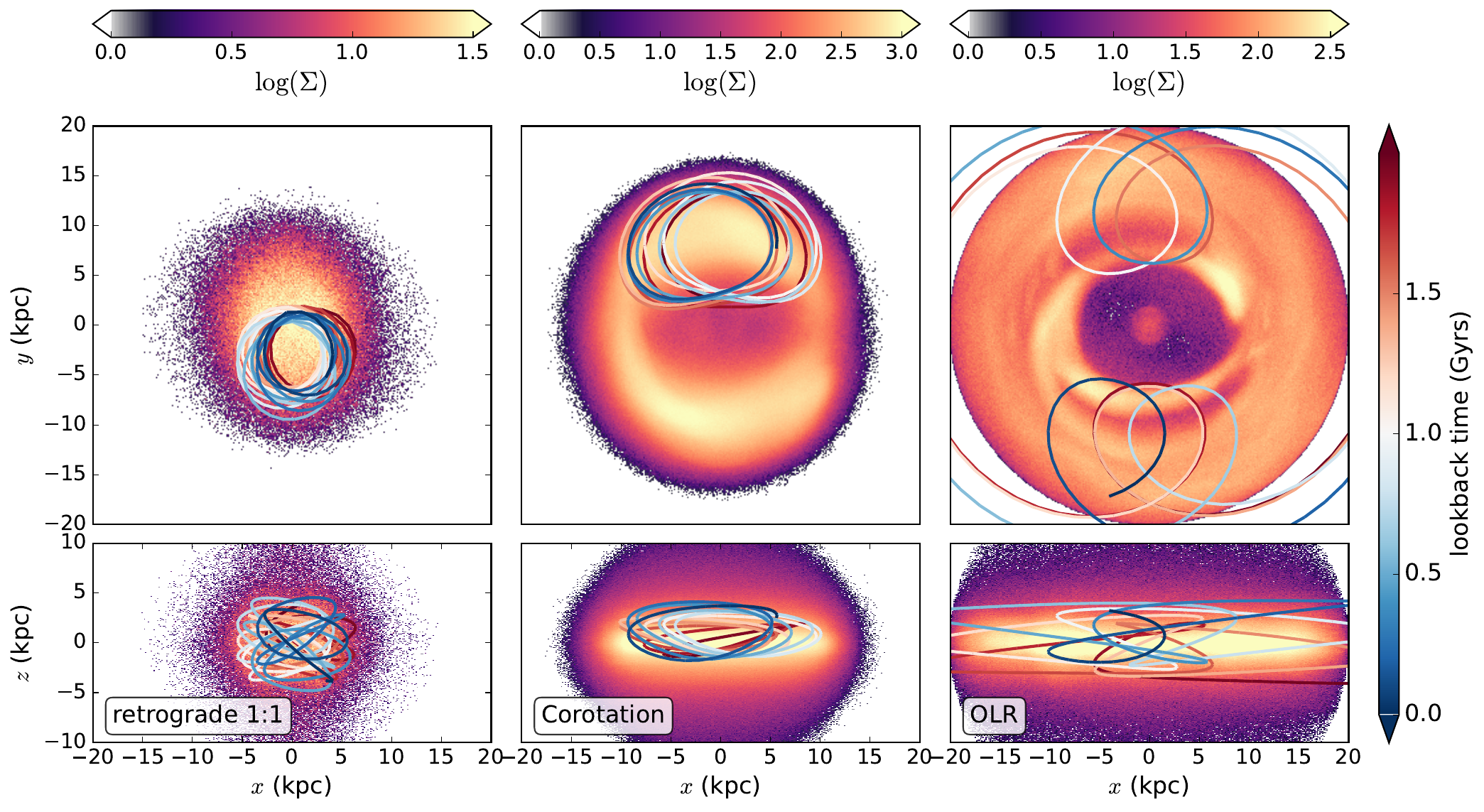}
    \label{fig:sub2}
    \caption{The  face-on (top panels) and edge on (bottom panels) density distribution of all stars within 20\,kpc in the retrograde 1:1 resonance (left), corotation resonance (middle) and OLR (right). Overplotted are examples of orbits from each family, in the reference frame rotating with the bar, with colours corresponding to lookback time.  In all panels the bar is located along the x-axis. We see that the 1:1 resonant stars are concentrated in the central regions, while the corotation stars have a peak in density at larger radii.}
    \label{fig:orbits}
\end{figure*}

\begin{figure*}
    \begin{subfigure}{\textwidth}
        \centering
        \centerline{\includegraphics[width=1.1\textwidth]{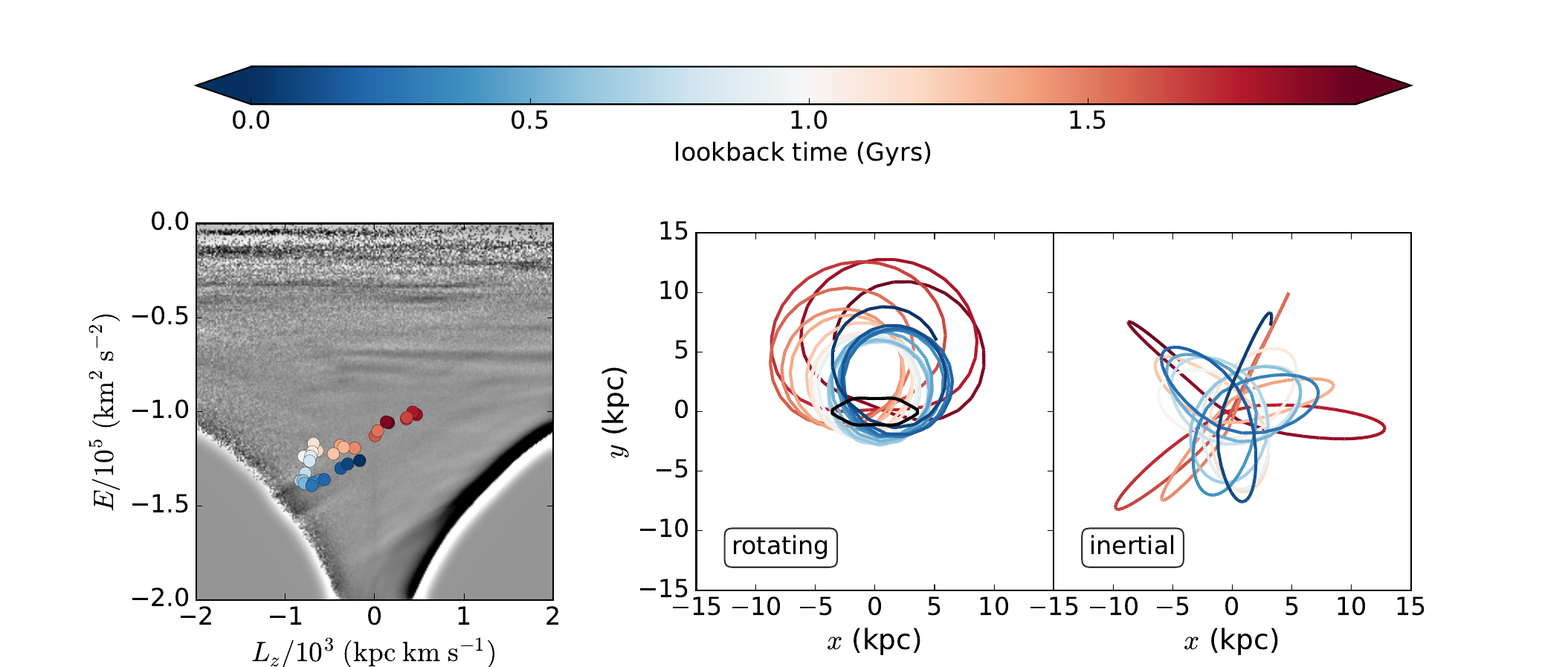}}
    \end{subfigure}
    \caption{An example of an orbit that changes between being prograde and retrograde and in doing so changes its orbital class from corotation to the retrograde 1:1 resonance. The left panel shows this in $E-L_z$ space with points taken from 40 different snapshots coloured by their lookback time, as shown in the colourbar at the top. The middle panel shows the orbit in face-on spatial coordinates in the reference frame rotating with the bar. The line is coloured by the lookback time of the star in the same manner as the points in the left panel. The right panel shows the orbit in the inertial frame, coloured in the same manner. The orbit starts out as prograde and becomes retrograde at a lookback time of around 1.4 Gyrs.}
    \label{fig:changing_orbit}
\end{figure*}

\section{Resonances in the stellar halo}
\label{sec:Results}

We start by exploring the overdensities in the distribution of all stars (i.e., both accreted and in-situ) in the $E-L_z$ plane, before moving on to explore the accreted stellar halo in more detail (\autoref{sec:e_lz}). We connect the overdensities seen in $E-L_z$ to the underlying bar-induced resonant orbital structure (\autoref{sec:config_space}), exploring how these orbits are distributed in configuration space and how they change over time. We then explore the chemical patterns in $E-L_z$ and how these relate to the resonant ridges (\autoref{sec:chemistry}).

\subsection{Overdensities in $E-L_z$} \label{sec:e_lz}

In \figref{fig:all summary} we show all stars (i.e. both in-situ and accreted) in the $E-L_z$ plane in the $R$ = 0\,kpc region (top panels) and in a "solar neighbourhood"-like region at $R$ = 8\,kpc (bottom panels). The first column shows the distribution of stars in this plane with an unsharp-masking filter applied to make substructures in this space more apparent, whereas the second column shows the distribution of stars coloured by logarithmic density, without an unsharp mask. The panels in the third column are colour-coded by the average stellar metallicity in each bin, while the rightmost panels are colour-coded by the average age of stars.

\enlargethispage{\baselineskip}

The arrow in the leftmost panels points towards an overdensity, or ridge, which is clearly seen in the retrograde part of the distribution, starting at $\mathrm{E \approx -1.5 \times 10^5 \; km^2 \; s^{-2}}$ and $\mathrm{L_z \approx -1 \times 10^3 \; kpc \; km \; s^{-1}}$ and moving in a diagonal line towards higher $E$ and $L_z$\footnote{For a comparison of the $E-L_z$ distribution between halo 18 in Auriga Superstars and fiducial Auriga, see Appendix A.}. As the ridge becomes more prograde, it becomes more horizontal. This ridge-like feature is also clearly distinguishable in both metallicity and age. Interestingly, its stellar populations are both younger and more metal-rich than the surrounding regions of $E-L_z$. 

As mentioned in \autoref{sec:intro}, overdensities or substructures in $E-L_z$ space of the stellar halo are frequently associated with remnants from past mergers. To explore the accreted population more explicitly, we plot the distribution of ex-situ stars in $E-L_z$ in Fig. \ref{fig:e-lz_resonanece}. 
The top panels show the region around $R$ = 0\,kpc and the bottom panels at $R=8$\,kpc. The grey 2D histograms in all panels shows the logarithmic density of the stars. As can be seen in the left panels, the aforementioned prominent ridge is even more pronounced in the accreted population.
The red diagonal line in the left panels shows a line of constant Jacobi energy, $E_J$, which we will return to below.

To determine whether the ridge could be caused by resonances related to the bar, we perform a frequency analysis of the orbits in the simulation, as described in \autoref{sec:Simulation and Methods}.
In the right panels, we add insets that show histograms of $r_{\Omega}$ (see \equarefalt{eq:resonance_r}) for the accreted stars in the two regions of interest. In these insets, we highlight the range of $r_\Omega$ values that we use to select stars at the different resonances\footnote{Note that we use different ranges of $r_{\Omega}$ at the different resonances to ensure that at least 2000 star particles are selected. Increasing or decreasing the interval in $r_\Omega$ used to identify the resonances does not shift the location of the contours, but it does affect the size of the region in $E-L_z$ that is traced by the contours.}. We find prominent peaks at the corotation ($r_{\mathrm{\Omega}}$ = 0) and 1:1 resonance ($r_{\mathrm{\Omega}}$ = -1), with a small number of stars at the OLR ($r_{\mathrm{\Omega}}$ = -0.5). The contours in the right-hand panels indicate the distribution of stars at these different resonances: the OLR orbits are outlined in yellow, corotation orbits in purple, and orbits in the 1:1 resonance in red.  
 
We see that the prominent ridge-like structure is largely made up of stars trapped in the corotation and 1:1 resonances: the prograde part of the ridge corresponds to stars in corotation, while the retrograde part is made up of stars trapped at the 1:1 resonance. In the prograde part of the ridge, a less prominent arm protrudes to higher energies to the region of $E-L_z$ space traced by stars in the OLR. Comparing the prominence of resonant stars in the two regions studied here, we note that there is an increased contribution of 1:1 resonant stars in the $R=0$\,kpc region, as compared to the $R=8$\,kpc region. There is also a slight increase in the contribution of stars at the OLR in the $R=8$\,kpc region as compared to the central region.

We can further see in the left panels of \figref{fig:e-lz_resonanece} that the slope of the ridge in the section corresponding to the retrograde 1:1 resonance, coincides with the line of constant $E_J$, which corresponds to a slope equal to the pattern speed of the bar. This suggests that, were such a ridge to be detected in the Milky Way, it would provide an independent measurement of the Milky Way's bar pattern speed. In \autoref{sec:Discussion} we discuss in detail the origin of the alignment between the 1:1 ridge and the Jacobi constant.

\begin{figure*}
    \begin{subfigure}{\textwidth}
        \includegraphics[width=\textwidth]{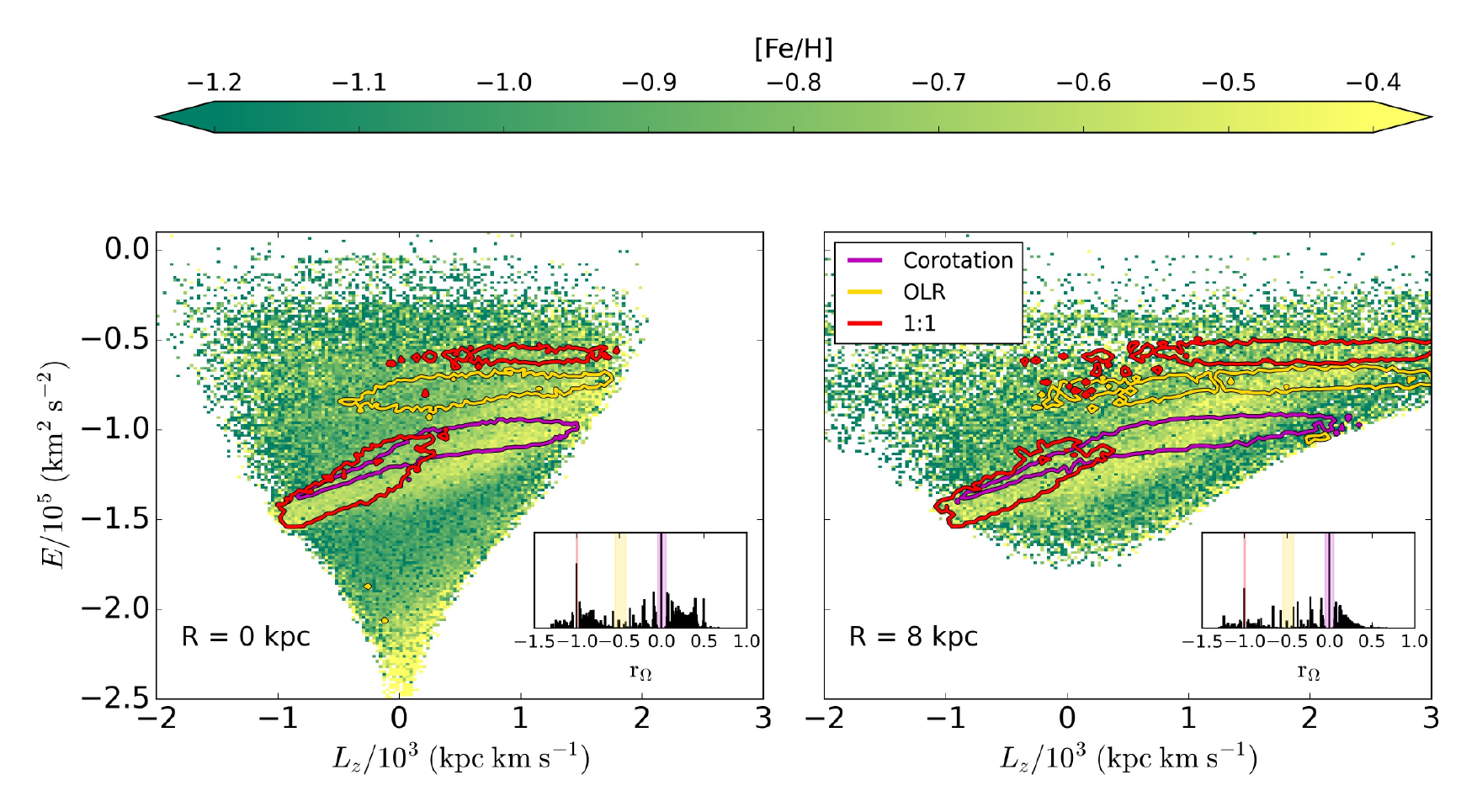}
    \end{subfigure}
    \caption{$E-L_z$ plots showing the accreted population in two spatial regions, a 4\,kpc sphere around the centre of the galaxy (left) and a 4\,kpc sphere around a "solar neighbourhood" (right). The bins are coloured by the mean [Fe/H] value in each bin. The differently coloured contours (90th percentile) indicate stars trapped in Corotation, OLR, and 1:1 resonances (CR: purple, OLR: yellow, 1:1: red), with an inset in each panel of the $\mathrm{r_{\Omega}}$ resonance value distribution for stars of each selection. Stars in the region of the ridge, which is traced by stars in the retrograde 1:1 resonance and corotation, have higher metallicities than stars in the surrounding $E-L_z$ space.}
    \label{fig:e-lz_metal}
\end{figure*}

To further explore the origin of the ridge and its relation to distinct accretion events, or to the bar, in Fig. \ref{fig:merger_density}, we show the distribution of stars associated with the four most massive accretion events in the galaxy in $E-L_z$ space. The contours in this figure show the location of resonant stars (selected from intervals in $r_\Omega$ as before)\footnote{The contours shown correspond to all resonant accreted stars within the 20\,kpc spatial selection that we use for this figure.}. We see that the ridge is prominent in at least two of the progenitors (M3 and M4), while it is also somewhat visible in M1 and M2, albeit not as strongly (in M1 and M2 the ridge is more clear in the prograde part). This lends further credence to the fact that the ridge is not caused by one individual accretion event, but is rather caused by a response of the stars to a global dynamical perturbation. We also note that the ridge is a feature that becomes stronger with time, after the formation of the bar, as illustrated in \figref{fig:ridge_evolution}.
In terms of absolute numbers, the dominant contribution to the ridge is from merger M4, which is the most massive of these four mergers (see Table \ref{tab:1}).
It is likely that the preferential trapping of stars from mergers M3 and M4 in the resonant ridge is related to the timing and the orbital configuration of the mergers: mergers with more radial orbits deposit stars at regions close to $L_z=0$, which, as we will see in \autoref{sec:Discussion}, allows these stars to be trapped by the corotation and 1:1 resonances. We will explore in detail the factors contributing to the trapping of stars in the resonances in future work (Gherghinescu et al. in prep.). 

\subsection{Connection to configuration space} \label{sec:config_space}

We have seen previously in \figref{fig:e-lz_resonanece} that the distribution of resonant stars differs between regions at $R=0$ and $R=8$\,kpc. These differences can be understood by examining the distribution of resonant orbits in configuration space, as shown in \figref{fig:orbits}. The figure shows the spatial surface density distribution of stars, both edge-on and face-on, at $z=0$, which are trapped (from left to right) at the retrograde 1:1, corotation and OLR resonances. Here we select all (in-situ and accreted) resonant stars within 20\,kpc of the centre of the halo. For illustrative purposes we overplot an example orbit for each of these resonances as a solid line, in the reference frame rotating with the bar, where the colour-coding corresponds to the lookback time (shown in the right colourbar).

We see in the left panel of the figure, that stars in the retrograde 1:1 resonance are most dense at the centre of the halo, with stars found out to a radius of $\sim$10\,kpc. Stars at this resonance have an almost spherical distribution, with no visible disc plane and a mean height above the plane of the disc of around 3.5 kpc. In the middle panel, we see that stars in the corotation resonance have the highest density in two regions perpendicular to the bar (in the vicinity of the L4 and L5 Lagrange points; e.g., \citealt{binney_2008}), at radii of $\sim8-10$\,kpc. Corotation orbits are more disc-like in their configuration than stars in the retrograde 1:1 resonance, but are also found at relatively large heights, away from the plane of the disc. Lastly, as seen in the right panel, OLR stars are found further out still, with densities highest at a radius of $\sim$15 kpc and extending beyond the 20 kpc region for which we have calculated the orbital frequencies. OLR stars are disc-like in configuration, with mean heights above and below the plane of the disc of around 2 kpc. 
Since the stars at the retrograde 1:1 resonance are most dense at the centre, and their density decreases with radius, as we probe regions out to larger radii, the contribution from retrograde 1:1 resonant stars decreases, while the contribution of stars in the corotation resonance and the OLR increases. This explains why the retrograde part of the ridge is seen more prominently in the top panels of Fig \ref{fig:e-lz_resonanece}, while the contribution from the CR and OLR increases in the bottom panels.

The regions where these orbits are found are likely to vary compared to the Milky Way, since this simulation is not an exact analogue of our Galaxy. As mentioned in \autoref{sec:Simulation and Methods}, while the length of the bar is similar between estimates of the Milky Way and this simulated halo, the corotation radius is different; this is approximately 10\,kpc in this simulation, which is larger than some of the most recent estimates of the corotation radius for the Milky Way \citep{bland_hawthorn_2016,sanders_2019,chiba_2021}. The difference in the spatial distribution of stars on these resonant orbits affects the visibility of overdensities caused by these resonant stars in $E-L_z$ space. In \figref{fig:other_sun} in appendix C, we show an example of how the ridge would look if we selected a region at a larger radius of 11.5\,kpc.

\subsubsection{Migration between resonant families}

When investigating the orbital properties of resonant stars, we find a subset of stars that are able to `migrate' between different resonant families. This is particularly prevalent in terms of migration between the corotation and retrograde 1:1 resonances, which are continuous and adjacent to each other in $E-L_z$ space. An example of such an orbit is shown in \figref{fig:changing_orbit}. In the left panel of the figure we show the position of the star along its orbit in $E-L_z$ space. The colours of the points denote the lookback time of the snapshot, as given in the colourbar at the top of the figure. The middle and right panels show the star's orbit over time in a rotating and inertial frame of reference, respectively, with the colour of the line also denoting the lookback time. Initially, the star is found in the region of $E-L_z$ space associated with corotation (see \figref{fig:e-lz_resonanece}) and its orbit is typical of corotation orbits (see e.g., \figref{fig:orbits}). At a lookback time of around 1.4 Gyrs, the star changes direction, from prograde to retrograde. It is now found in the region of $E-L_z$ space associated with stars in 1:1 resonance, and its orbit in configuration space now looks like a typical 1:1 resonant orbit (as seen most clearly in the rotating frame of reference). 

\begin{figure*}
    \centering
    \begin{subfigure}{\textwidth}
        \centering
        \includegraphics[width=.9\textwidth]{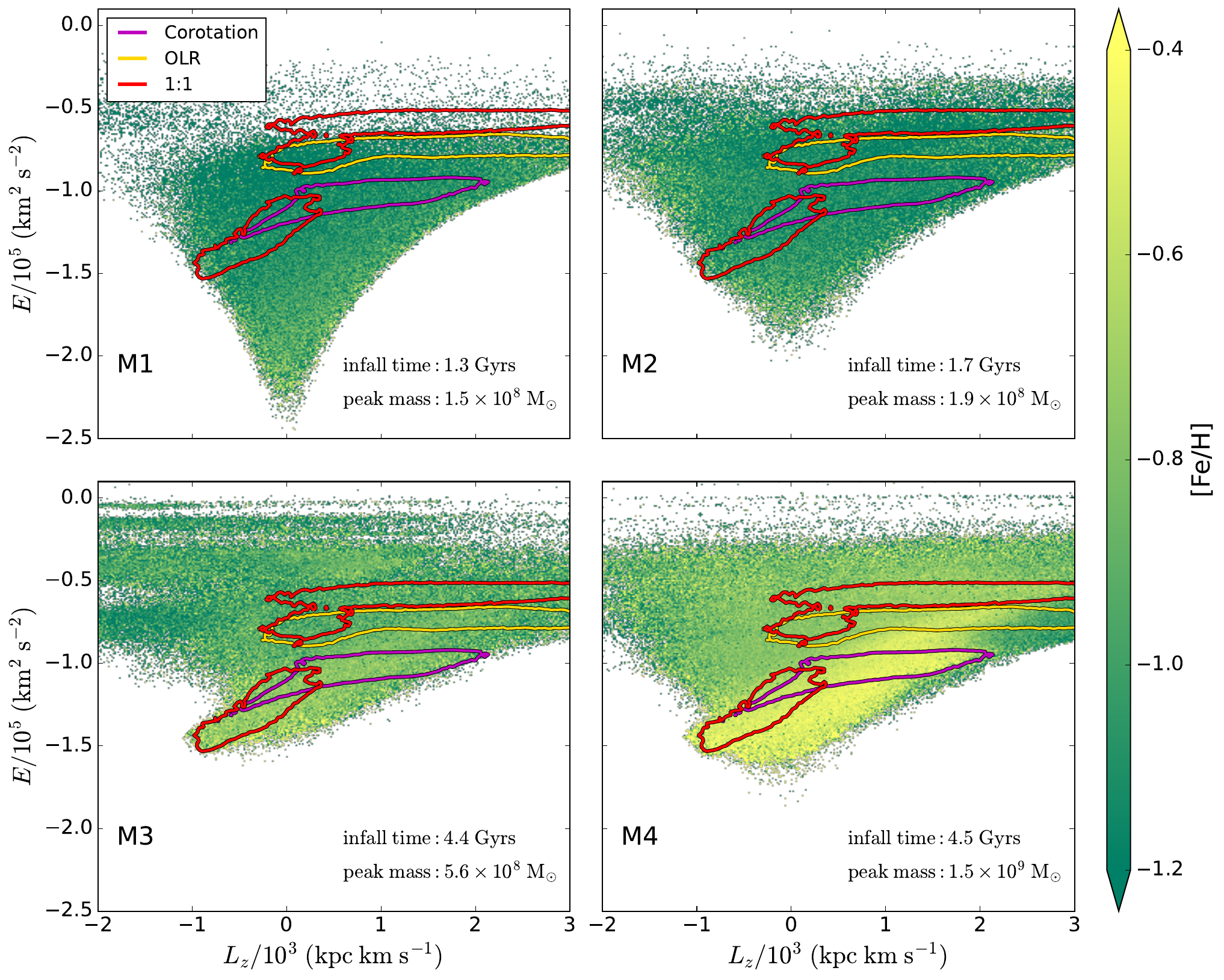}
    \end{subfigure}
    \caption{The four panels show stars from the four main accreted progenitors in the simulation coloured by their mean metallicity in each bin in $E-L_z$ space. All stars from these progenitors within 20 kpc are selected. Differently coloured contours (90th percentile) show the distribution of resonant stars for all accreted stars (CR: purple, OLR: yellow, 1:1: red). The mergers are in order of their infall times (left to right, top to bottom). Mergers M1 and M2, which merged several Gyrs before bar formation, show only slight trends in metallicity, with higher metallicity at lower energy, but the ridge is not apparent in metallicity. Stars from mergers M3 and M4, which merged just before bar formation, have higher metallicities than stars from the two earlier mergers and have higher metallicities in the region of the ridge compared to stars at larger energies. In combination with the densities shown in \figref{fig:merger_density} this suggests that the distinct metallicity of the ridge as seen in \figref{fig:e-lz_metal} is caused by stars from these two later mergers.}
    \label{fig:merger_metal}
\end{figure*}

We find  multiple orbits that behave in similar ways and move similar distances in $E-L_z$ space. This suggests that this is a potential pathway for stars to go from being prograde to retrograde, and vice versa, without the need of a merger -- i.e., these stars are made retrograde because of internal bar-driven resonant processes. 
We will further quantify the exact extent to which this process happens by using both cosmological and controlled simulations in future work. 

\subsection{Chemistry of resonant ridges} \label{sec:chemistry}

We have seen so far that bar-induced resonances can create overdensities in $E-L_z$ space. Additionally, as shown in the third and fourth panels of \figref{fig:all summary}, the ridge is distinct from its surroundings in terms of its metallicity and ages. Since chemical tagging is one of the ways employed to distinguish accreted from in-situ populations \citep{gilmore_1989, freeman_2002, hawkins_2015}, in combination with stellar dynamics, it is important to understand whether bar-induced resonances can also create such chemically distinct substructures in the halo. 

In \figref{fig:e-lz_metal} we show the accreted population in $E-L_z$ space in bins coloured by the mean metallicity of stars. We do this in the two regions we explore, i.e., the central region (left) and the "solar neighbourhood" region (right). The contours in this figure show the location of resonant stars, as previously. We can see that the stars inside the prominent ridge are more metal-rich than stars outside the ridge (both above and below the ridge), with a difference in [Fe/H] of around 0.5\,dex. Thus, the ridge, which is a dynamical feature formed because of the influence of the bar, has a population of accreted stars with a chemistry that is distinct from the accreted population outside the ridge.

To explore how this distinct chemistry arises, and how this might relate to the accretion events in the galaxy -- in addition to the bar-induced resonances -- in \figref{fig:merger_metal} we explore the metallicity in $E-L_z$ of the four most massive mergers within 20\,kpc of the centre.
We see that mergers M1 and M2 have quite uniform metallicities, which are overall lower than for M3 and M4 (due to its higher mass, M4 has the highest metallicity). On the other hand, mergers M3 and M4 have a gradient in metallicity, in which the metallicity is higher at lower energies. This is due to the fact that these progenitors had negative radial metallicity gradients at infall, with higher metallicity at the centre, which is mapped into a negative metallicity gradient in energy, since the outer parts of the galaxy are stripped first \citep{orkney_2023}. This is also found in observations of dwarf galaxies in the local group (e.g. \citealt{taibi_2022}). In GES-like progenitors in the Auriga simulations, this initial metallicity gradient, while blurred, persists until $z=0$, with higher metallicities at lower energies \citep{carrillo_2025}.

In addition to this overall negative metallicity gradient in energy, we see that the M3 and M4 remnants have a metallicity distribution with a diagonal feature, which follows the 1:1 resonance. As we will discuss in the next section, this is due to how stars are scattered and trapped in the bar-induced 1:1 resonance, with stars from regions of low angular momentum (i.e. $L_z\sim$0) moving towards the retrograde part of the distribution (i.e. $L_z <$ 0) and towards lower energies, i.e., scattering along the 1:1 resonance line.
Since most of the high-metallicity stars in the ridge originate from M4 (since it is the most massive of the four mergers), this causes a ridge of higher metallicity, while at lower energies the M1 and M2 progenitors contribute more (as they sink deeper into the potential well, since they are earlier merger events).

\section{Orbital scattering at the resonances}
\label{sec:Discussion}

\begin{figure}
    \centering
    \includegraphics[width=\columnwidth]{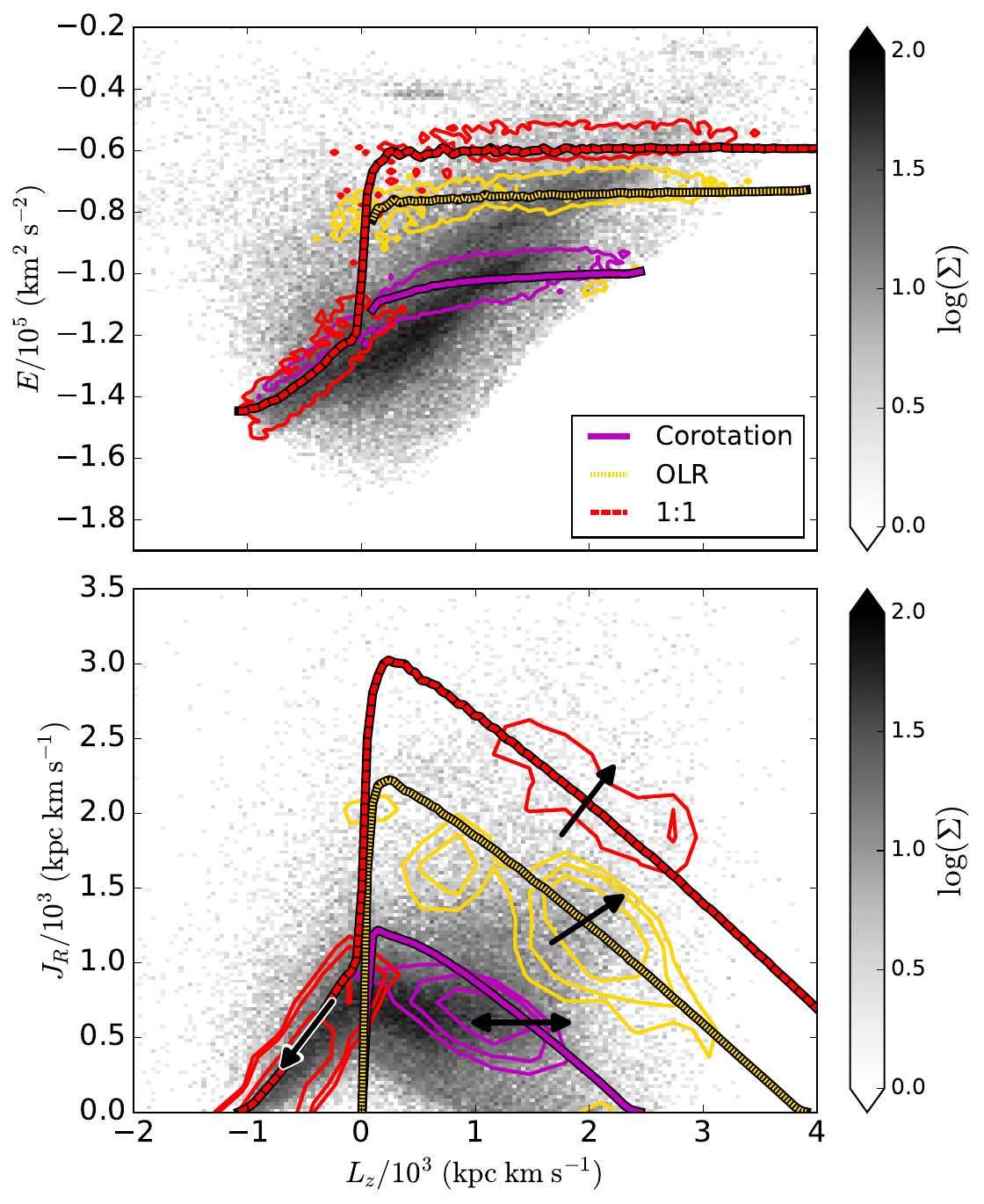}
    \caption{The top panel shows in the background the distribution of all stars in the "solar neighbourhood" region in $E-L_z$ space with contours showing the 90th percentile of distributions of stars in corotation (purple), OLR (yellow), and 1:1 resonance (red). In addition to the contours, three sets of lines (with the colours the same as for the contours) show the locations of the axisymmetric resonance lines (ARLs), which are calculated from the axisymmetric potential fit to the simulation. The bottom panel shows the same selection of accreted stars in action space ($\mathrm{J_R - J_{\phi}}$) as the grey background distribution. The contours here show the distribution of stars in the corotation, OLR, and 1:1 resonances (66th, 50th, 33rd percentiles). The three sets of lines show the ARLs in this plane as in the upper panel. Additionally, the arrows demonstrate the slopes along which individual stars should move at the different resonances (see text). For the 1:1 retrograde stars this slope aligns with the slope of the ARL.}
    \label{fig:theory}
\end{figure}

One of the findings of our study is that the retrograde part of the 1:1 ridge in $E-L_z$ space appears to have a slope equal to that of a line of constant Jacobi Energy $E_J = E-\Omega_{\mathrm{bar}}L_z$, i.e. with a slope equal to $\Omega_{\mathrm{bar}}$. Therefore, if the 1:1 retrograde ridge were to be detected in the Milky Way, it could give us an independent measurement of the bar pattern speed, which is still rather uncertain, with studies placing it between $\sim$25 and 50 $\mathrm{km \; s^{-1} \; kpc^{-1}}$ (e.g., \citealt{sanders_2019, fragkoudi_2019, binney_2020, clarke_2022, zhang_2024, horta_2025}). We also showed in the previous section that the 1:1 retrograde ridge has a higher metallicity than its surroundings, suggesting that stars scatter from regions of higher energy (and therefore metallicity) towards more retrograde angular momentum and lower energies. 

In this section, we investigate how stars scatter at the 1:1 resonance, which can shed light on why the slope of the retrograde 1:1 resonant ridge is equal to $\Omega_{\mathrm{bar}}$. First, we start by exploring the slope along which stars are scattered at the 1:1 resonance in $E-L_z$ and action space (\autoref{sec:slopescatter}). Subsequently, we explore whether stars gain or lose angular momentum during this scattering process (i.e. determine the direction of scattering; \autoref{sec:dirscatter}). We then discuss the overall changes in $L_z$ and $J_R$ at the resonances in our simulation (\autoref{sec:changelzJR}), before commenting on the role of the scattering of stars at the resonances on the chemistry of the ridge (\autoref{sec:action_metal}).

\subsection{Slope of scattering in action space}
\label{sec:slopescatter}
We start by exploring the axisymmetric actions, i.e. the actions calculated for an axisymmetric approximation of the potential. We also look at the axisymmetric resonance lines (ARLs), as used by \citet{trick_2021}, which are obtained from the frequencies associated with a star's axisymmetric actions, which we label as $\Omega_{\mathrm{axi},i}$, where $i$ denotes \{$R$, $\phi$\}\footnote{To obtain the ARLs, we select stars which are confined close to the plane, with $J_z < 40 \mathrm{\; kpc \; km \; s^{-1}}$, whose orbits satisfy the resonance condition in Equation \eqref{eq:resonance}. See \citealt{trick_2021} for more details.}. These lines can give us a sense of where in $E-L_z$ or action space the resonances would lie in an axisymmetric potential. 
In \figref{fig:theory} we show the distribution of accreted stars in the "solar neighbourhood" region as the grey background in both the $E-L_z$ plane (top panel) and action space $J_{\phi}-J_{R}$ (bottom panel). On top of these we plot contours outlining the distribution of resonant stars from the simulation, as well as the ARLs for the CR, 1:1 and OLR as differently coloured and dotted lines. 


One important difference between the $J_R-J_{\phi}$ and $E-L_z$ spaces, is the way stars will move in these at the resonances: in $E-L_z$ space, stars move along lines of constant $E_J$, with a slope equal to $\Omega_{\mathrm{bar}}$. On the other hand, in action space, stars move along different slopes, depending on their location in this plane, i.e. where they are related to the different resonances \citep{binney_2008, sellwood_2002}. This can be understood by examining the fact that the Jacobi Energy, $E_J$, is constant:

\begin{equation}
    E_J = E - \Omega_b L_z = \mathrm{constant}
    \label{eq:action_1}
\end{equation}

\noindent which leads to,

\begin{equation}
    \mathrm{d}E = \Omega_{\mathrm{bar}}\mathrm{d}L_z.
    \label{eq:action_2}
\end{equation}

\noindent Therefore, when $E_J$ is  an integral of motion, stars move along slopes of $\Omega_{\mathrm{bar}}$ in the $E-L_z$ plane.
To understand how stars move in action space, we can expand out $\mathrm{d}E$ into partial derivatives of the actions, assuming that these are independent of each other (see also \citealt{sellwood_2002}). 
We further assume a negligible change in the vertical action $J_z$, i.e. $\partial E/\partial J_z = 0$. Thus, the previous equation becomes:

\begin{equation}
    \Omega_\mathrm{b} \;\mathrm{d}L_z = \frac{\partial E}{\partial J_R} \mathrm{d}J_R + \frac{\partial E}{\partial J_\phi} \mathrm{d}J_\phi.
    \label{eq:action_3}
\end{equation}

\noindent 

\noindent In the axisymmetric approximation, $\mathrm{d}E/\mathrm{d}J_R =  \Omega_{\mathrm{axi},R}$ and $\mathrm{d}E/\mathrm{d}J_{\phi} =  \Omega_{\mathrm{axi},\phi}$ \citep{binney_2008}. Substituting the orbital frequencies into the previous equation, we get,

\begin{equation}
    \Omega_\mathrm{b} \;\mathrm{d}L_z = \Omega_{\mathrm{axi},R} \;\mathrm{d}J_R + \Omega_{\mathrm{axi},\phi} \; \mathrm{d}J_{\phi}.
    \label{eq:action_4}
\end{equation}

\noindent Rearranging this equation and using $J_\phi = L_z$, gives:

\begin{equation}
    \mathrm{d}J_R = \frac{\Omega_{\mathrm{bar}} - \Omega_{\mathrm{axi},\phi}}{\Omega_{\mathrm{axi},R}} \;\mathrm{d}L_z.
    \label{eq:action_5}
\end{equation}

\noindent Finally, we can now use the definitions in Equation \eqref{eq:resonance} to get a relation for how stars move in the $J_R-J_\phi$ plane:

\begin{equation}
    \Delta J_R = \frac{l}{m} \Delta L_z.
    \label{eq:action_6}
\end{equation}

Here $\Delta J_R$ is the change in the radial action, and $\Delta L_z$ the change in angular momentum (or the azimuthal action $J_{\phi}$). We thus expect stars to move in the $J_R-J_\phi$ plane along slopes of $l/m$, where $l$ and $m$ are integers identifying the resonance (see also \citealt{sellwood_2010, trick_2021}). These slopes are shown in the bottom panel of \figref{fig:theory} as black arrows.

By examining the $J_R - J_\phi$ space, and the slopes of the arrows (i.e. the way in which stars scatter at the resonances), we can begin to see why there is a strong ridge forming at the 1:1 resonance, with a slope equal to 1; for the retrograde 1:1 resonance, the scattering vector coincides with the slope of the ARL. Thus, stars in the retrograde 1:1 resonance experience strong scattering since they remain on the resonance as they scatter in $J_R - J_\phi$ space\footnote{A similar process has been suggested to occur for stars at the ILR by \citet{sellwood_2010}. For these stars, the scattering vector is also aligned with the location of the ILR in $J_R - J_\phi$ space, which allows for stars to remain on resonance as they are scattered.}. This is what drives the strong ridge feature at the retrograde 1:1 resonance. This is also reflected in the $E-L_z$ space: the slope of the ARL at the 1:1 retrograde resonance is aligned with the slope of $E_J$, which corresponds to the slope along which stars scatter in this space. Therefore, stars at the 1:1 resonance scatter along $E_J$, staying on resonance as they scatter. On the other hand, stars at the CR, OLR, and prograde 1:1 resonances, move away from their ARLs when they are scattered at these resonances. 

\subsection{Direction of scattering}
\label{sec:dirscatter}

In the above, we discussed the slope with which stars scatter at the 1:1 resonance, and we showed that in $J_R - J_{\phi}$ space the scattering of stars at the 1:1 resonance has the same slope as the ARL. However, determining the direction of angular momentum transport at this resonance (i.e., the direction in which the arrow points) is also relevant for understanding the shape of the 1:1 ridge: when stars scatter at the retrograde 1:1 resonance, do they gain, or lose angular momentum? The change in angular momentum at the ILR, CR and OLR resonances is discussed in \citet{lynden_bell_1972}; here we follow their steps to understand whether one would expect stars at the retrograde 1:1 resonance to lose or gain angular momentum. \citet{lynden_bell_1972} show that the change in angular momentum at the resonances is given by:

\begin{equation}
    \dot{H}_{lm} = -\frac{1}{8\pi}\iint_{0}^{\infty} m(l\frac{\partial F}{\partial J_R} + m\frac{\partial F}{\partial J_\phi}) |\psi_{lm}|^2 \delta(l\Omega_R+m\Omega_\phi+\omega) \mathrm{d}J_R \mathrm{d}J_{\phi}.
    \label{eq:ang_m exchange}
\end{equation}

\noindent where $\dot{H}$ is the change in angular momentum. The right-hand side of the equation is an integral over the radial and azimuthal actions of the distribution function, $F$, which itself is a function of $J_R$ and $J_\phi$.
As previously, $l$ and $m$ refer to the integer values describing the resonances. The delta function inside the integral stipulates that angular momentum exchange occurs at the resonances, while $\psi_{lm}$ are Fourier coefficients describing the perturbing potential. 

In order to understand whether angular momentum is gained or lost at a given resonance, we need to determine the sign of the following part of the equation, 

\begin{equation}
    m(l\frac{\partial F}{\partial J_R} + m\frac{\partial F}{\partial J_\phi}). 
\end{equation}

If this expression is positive, then $\dot H$ is negative, and stars lose angular momentum at a resonance. If it is negative, they gain angular momentum. In the case of the 1:1 resonance, both $m$ and $l$ are equal to 1. The overall sign of the expression depends on the distribution function, $F$, and its partial derivatives. Generally, as assumed by \citet{lynden_bell_1972}, the partial derivative $\partial F/ \partial J_R$ is negative if the distribution function decreases with increasing epicylic amplitude, i.e. we get fewer orbits as they become less circular. $\partial F / \partial J_\phi$ also tends to be negative at positive $J_\phi$ due to the radial decrease in surface brightness in galaxies. For $J_\phi<0$, $\partial F / \partial J_\phi$ becomes positive, as the distribution function decreases for larger absolute values of $J_\phi$. This is the case for reasonable distribution functions, as well as for the case of a simple spherical halo\footnote{It is straightforward to create a spherical distribution function that satisfies this condition. All that is needed is a slight radial velocity anisotropy. This is the case for the halo of our simulation, and is also the case for the stellar halo of the Milky Way \citep{deason_2013}, as well as in other cosmological simulations \citep{sales_2007, navarro_2010}.}. Thus, as long as $|\partial F / \partial J_\phi| > |\partial F / \partial J_R|$, which is the case for most standard distribution functions, stars will lose angular momentum (i.e., the angular momentum becomes more negative) at the retrograde 1:1 resonance. As a consequence, stars will in general be scattered from $L_z\sim 0$ to more retrograde orbits. 

\subsection{$L_z$ and $J_R$ changes at the resonances}
\label{sec:changelzJR}

\begin{figure}
    \begin{subfigure}{\columnwidth}
        \includegraphics[width=1.\columnwidth]{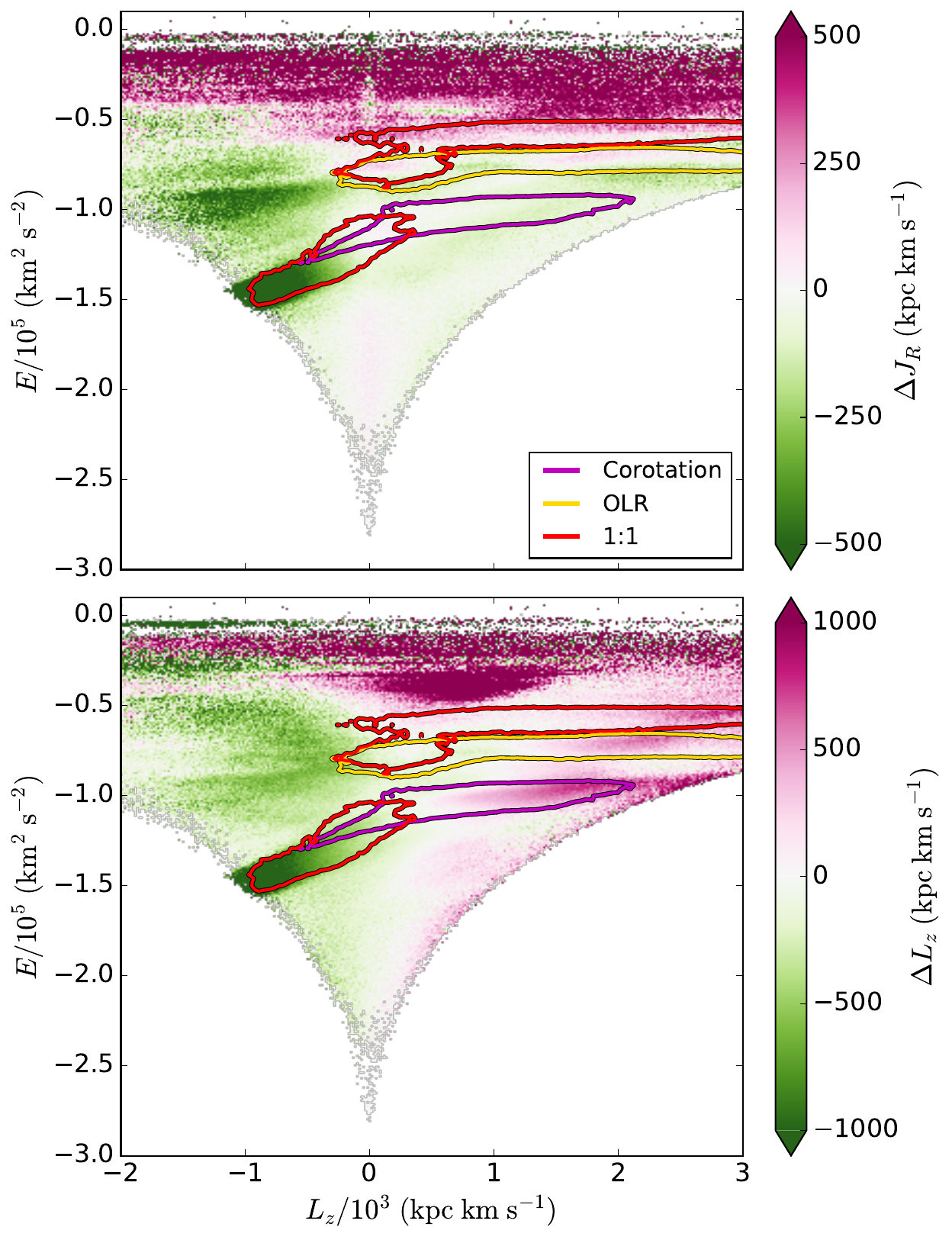}
    \end{subfigure}
    \caption{The top panel shows a plot of $E-L_z$ space with bins coloured by the average change in the radial action $J_R$ of accreted stars in each bin, for all accreted stars within a radius of 20 kpc from the centre. The change in $J_R$ is between the snapshot at redshift 0 and an earlier snapshot at a redshift of 0.79, corresponding to a lookback time of 7 Gyrs. Green denotes a loss in $J_R$, which means that an orbit has become more circular, and pink bins have seen an increase in $J_R$. Contours (90th percentile) show the distribution of stars in corotation (purple) and 1:1 resonance (red), as well as the OLR (yellow). From this plot we can see that stars in the retrograde 1:1 resonance have significantly lower $J_R$ now than they initially started out with. This agrees with the scattering of stars proposed in the lower panel of \figref{fig:theory}. Additionally, the $E-L_z$ plot in the bottom panel shows the change in $L_z$ for the same selection of accreted stars over the same time period. Stars at the retrograde 1:1 resonance have gained a significant amount of negative angular momentum, whereas stars at corotation, the OLR, and the prograde 1:1 resonance have slightly gained positive angular momentum.}
    \label{fig:Jr_change}
\end{figure}

Now that we have an expectation about how stars will scatter at the 1:1 retrograde resonance -- i.e., we know that they will move along a line with slope 1, towards more negative angular momenta -- we can explore how the actions of stars trapped in different resonances change in our simulation. We explore this by comparing the actions of stars at present day to an earlier snapshot (at a lookback time of 7\,Gyrs, $z=0.79$) which corresponds to a time just after bar formation. This is shown in \figref{fig:Jr_change}, where we show all accreted stars with $r<20\,\mathrm{kpc}$ in the $E-L_z$ plane with bins coloured by the mean change in the stars' actions between the present day and the $z=0.79$ snapshot. The top panel shows the change in $J_R$, i.e. $\Delta J_R = J_{R, \, z=0} - J_{R, \, z=0.79}$ 
while in the bottom panel we show the change in $L_z$, i.e. $\Delta L_z = L_{z, \, z=0} - L_{z, \, z=0.79}$. The contours in this figure show the location of resonant stars, as previously. 

Focusing first on the corotation resonance, we see that there is no significant change in $J_R$, while $J_{\phi}$ is increased over the bar's lifetime; this is in agreement with what we expect for stars at the corotation resonance \citep{sellwood_2002}. On the other hand, as expected from the discussion above, the retrograde 1:1 resonance shows a decrease in $J_R$ as well as a decrease in $L_z$, i.e. stars at the retrograde 1:1 resonance become more circular over time with a more pronounced retrograde motion. Thus, stars that are currently in the retrograde 1:1 resonance were scattered there from regions in $E-L_z$ space with less negative angular momentum and with higher $J_R$ (and higher $E$). The ridge is drawing stars from higher density regions of $E-L_z$ and depositing them at more retrograde locations, increasing the density of stars there compared to surrounding regions in $E-L_z$ space, thus creating a prominent ridge. 

\subsection{Metallicity}
\label{sec:action_metal}

The scattering of stars at the resonances can also help in understanding the distinct metallicity of the ridge seen in Figs. \ref{fig:all summary} and \ref{fig:e-lz_metal}. High-metallicity stars in the ridge originate overwhelmingly from mergers M3 and M4, which are the most recent and massive mergers. As discussed above, these mergers had progenitors with radial metallicity gradients which translate into metallicity gradients in energy (e.g. \citealt{carrillo_2025}), with higher metallicity stars populating lower energies. The scattering of stars both towards the 1:1 retrograde and corotation resonances, creates a diagonal ridge with high metallicity stars, visible in both Fig. \ref{fig:merger_metal} and Fig. \ref{fig:e-lz_metal}. The high metallicity ridge seen in Fig. \ref{fig:e-lz_metal} is accentuated by the fact that the surrounding stars have lower metallicities: at higher energies this is caused due to the metallicity gradient in energy, and at lower energies this is because mergers M1 and M2 contribute more, since they sink lower into the host potential, and have lower metallicities. We hypothesise that higher metallicity stars may also be more prominent in the ridges because they tend to have lower $J_z$, which makes them more susceptible to being trapped by bar-induced resonances (see also \citealt{dillamore_2024}). This is also commonly seen for disc stellar populations, in which colder populations are more efficiently trapped by the bar resonances (see e.g., \citealt{fragkoudi_2017, debattista_2017}). 

A similar process leads to the distinct metallicity of the ridge in Fig. \ref{fig:all summary} which includes both in-situ and accreted stars, in which the ridge has a higher metallicity than the surrounding phase space. In-situ, high metallicity stars are found mainly in the prograde region close to the maximal $L_z$. However, due to past mergers some in-situ stars have had their orbits altered to lower $L_z$ -- for example, in the Milky Way, the `splash' is a proposed in-situ population heated by a merger \citep{belokurov_2020}. As seen in Fig. \ref{fig:all summary}, in our simulation we find a metallicity and age gradient with respect to $L_z$, i.e. stars that are retrograde have on average lower metallicities and older ages than prograde stars. When stars get trapped at the retrograde 1:1 resonance, they are scattered from regions with $L_z \approx 0$, towards more retrograde values of $L_z$ and lower energies. Therefore, the stars that are scattered by the resonance have higher metallicities than the population that is already present in the surrounding retrograde phase space, leading to ridge of resonant stars with higher metallicities.

While we will explore in future work the various mechanisms that can give rise to the increased metallicity in the ridge, it is clear that this is likely due to a combination of dynamical effects along with the accretion history of the halo, which determines the distribution function of the halo at the time of bar formation. Thus, the details of the metallicity of the ridge might look different to those of the Milky Way, depending on the exact merger history of the Galaxy. Nevertheless, we find that bar resonances can cause distinct metallicity features in $E-L_z$ space; caution is therefore needed when associating chemically distinct populations with any single accretion event, since, at least in this simulation, such a feature is caused by the bar resonances and combines different accretion events.

\section{Conclusions}
\label{sec:Conclusions}

While bars have long been known to induce resonances in their surrounding dark matter haloes (e.g. \citealt{athanassoula_2002}), little is known about how they affect the \emph{stellar} haloes of their host galaxy. Recent studies of Gaia data and test particle simulations, have suggested that bar resonances can induce substructures in the stellar halo, visible in $E-L_z$ space \citep{dillamore_2023, dillamore_2024}. In this work, we use a high-resolution cosmological zoom-in simulation from the Auriga Superstars suite, to show that such bar-induced substructures are present in the accreted stellar halo of a Milky Way analogue. 
This is the first such detection in a cosmological simulation, which allows us to study bar-induced substructures along with the assembly history and chemical enrichment of the galaxy.


Our results can be summarised as follows: 

\begin{description}
\item[(i)] We find that the bar-induced resonances, such as the retrograde and prograde 1:1, corotation and Outer Lindblad Resonance, cause a number of overdensities or ridges in $E-L_z$ space. 

\item[(ii)] The most prominent of these overdensities is a ridge stretching from negative to positive angular momentum, and consists of a retrograde part composed of stars at the 1:1 resonance, while the prograde part is associated with stars at the corotation resonance. 

\item[(iii)] This overdensity is visible in both the central region of the galaxy as well as in a "solar neighbourhood"-like region. There are differences in appearance of the ridge in the two regions, caused by how the various resonant families populate configuration space.

\item[(iv)] Stars from several progenitors are found within the ridge, clearly indicating that this cannot be associated with a single accretion event.

\item[(v)] By exploring the orbits of resonant stars, we find that they can migrate between orbital families, switching between the corotation and retrograde 1:1 resonance and vice versa, thus changing between being prograde and retrograde. 

\item[(vi)] We explore the scattering of stars at the 1:1 retrograde resonance, finding that stars become more circularised with more negative angular momentum, i.e. they are scattered from regions of $L_z\sim 0$ towards more negative $L_z$ and lower energies. This is confirmed by tracing the evolution of the radial action and angular momentum of resonant 1:1 stars since bar formation, finding that their $J_R$ and $L_z$ decrease.

\item[(vii)] We find that the retrograde part of the ridge, which corresponds to the 1:1 resonance, has a slope equal to the bar pattern speed. We explain the reason for this by exploring the scattering of stars at the resonances in $E-L_z$ and action space. If such a ridge were to be identified in the Milky Way, it could provide an independent measure of the bar's pattern speed. 

\item[(iix)] The bar-induced ridge has a higher metallicity than the surrounding stars. This is caused by the preferential trapping of metal-rich stars in the resonances, which in turn is due to where these are deposited in phase space by the various accretion events taking place in the galaxy. 

\end{description}


Exploring the stellar halo of the Milky Way in spaces of integrals of motion, such as $E-L_z$, provides an opportunity to uncover the accretion history of our Galaxy. Our study suggests that this space can also be used  to uncover information about the Galaxy's bar and its coupling with the stellar halo. However, our results also suggest the need for caution when associating substructures in $E-L_z$ space -- even when combined with chemical information -- to distinct accretion events, since such features can in fact be caused by internal perturbations in the galaxy.

\section*{Acknowledgements}

TT acknowledges funding from the Science and Technology Facilities Council 2876865. 
FF and PG are supported by a UKRI Future Leaders Fellowship (grant no. MR/X033740/1). 
AF acknowledges support by a UKRI Future Leaders Fellowship (grant no MR/T042362/1) and a Swedish Wallenberg Academy Fellowship. 
AC and AD acknowledge support from the Science and Technology Facilities Council (STFC) [grant numbers ST/T000244/1 and ST/X001075/1] and the Leverhulme Trust. AD is supported by a Royal Society University Research Fellowship. 
RJJG acknowledges support from an STFC Ernest Rutherford Fellowship (ST/W003643/1).
FAG acknowledges support from the ANID BASAL project FB210003, from the ANID FONDECYT Regular grant 1251493 and from the HORIZON-MSCA-2021-SE-01 Research and Innovation Programme under the Marie Sklodowska-Curie grant agreement number 101086388.
FvdV is supported by a Royal Society University Research Fellowship (URF/R1/191703 and URF/R/241005).
RB is supported by the SNSF through the Ambizione Grant PZ00P2\textunderscore223532.
This work used the DiRAC@Durham facility managed by the Institute for Computational Cosmology on behalf of the STFC DiRAC HPC Facility (www.dirac.ac.uk). The equipment was funded by BEIS capital funding via STFC capital grants ST/P002293/1, ST/R002371/1 and ST/S002502/1, Durham University and STFC operations grant ST/R000832/1. DiRAC is part of the National e-Infrastructure.

\section*{Data Availability}

The data underlying this article will be shared on reasonable request to the corresponding author.



\bibliographystyle{mnras}
\bibliography{main} 




\clearpage
\appendix

\section{Fiducial Auriga resolution comparison}

\begin{figure}
\centering
    \begin{minipage}[c]{\columnwidth}
        \centering
        \includegraphics[width=\linewidth]{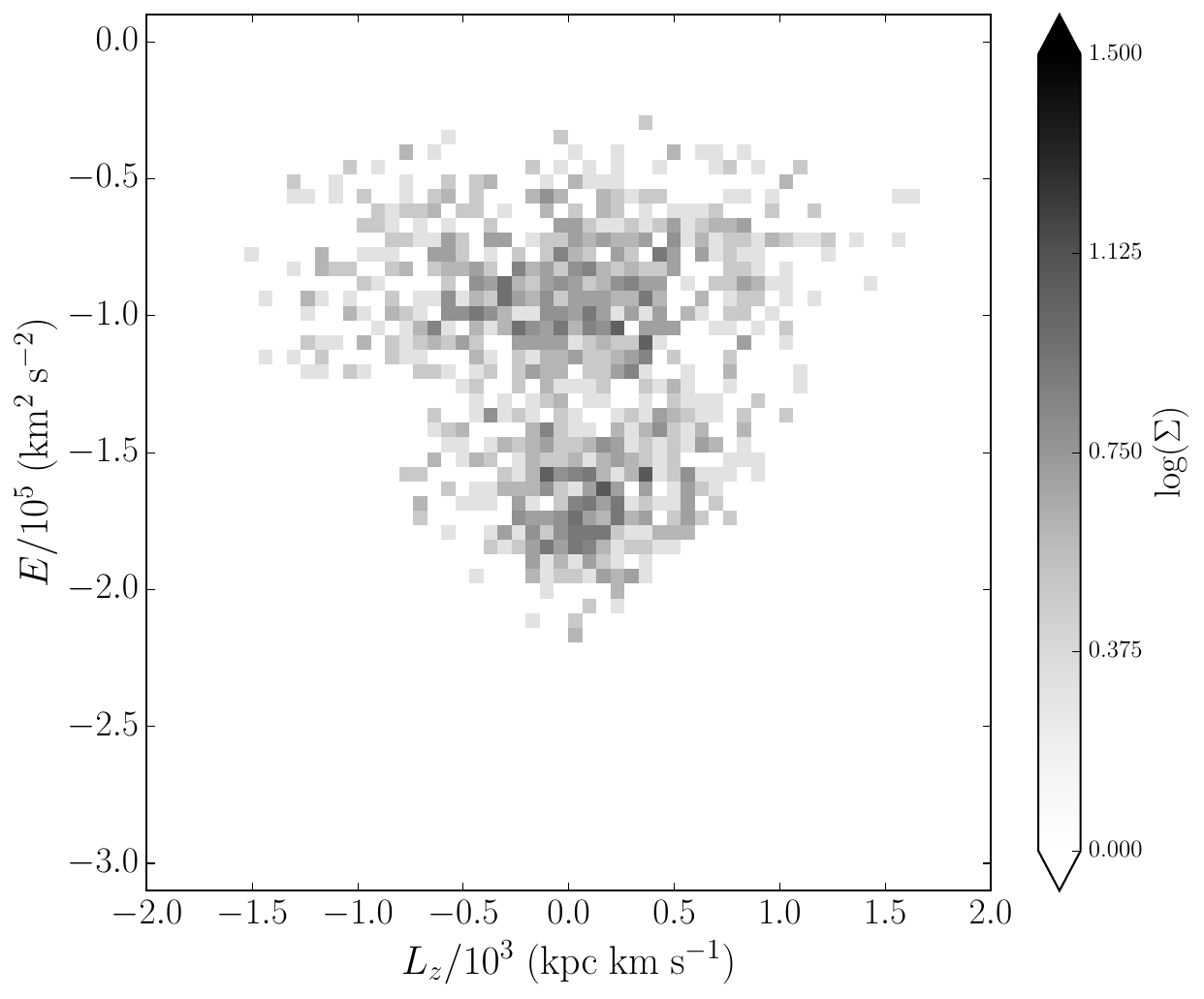}
        \vspace{0.5em} 
        \parbox[c]{\linewidth}{\centering \large R = 0 kpc}
        \label{fig:L4_centre}
    \end{minipage}
    \hfill
    \begin{minipage}[c]{\columnwidth}
        \centering
        \includegraphics[width=\linewidth]{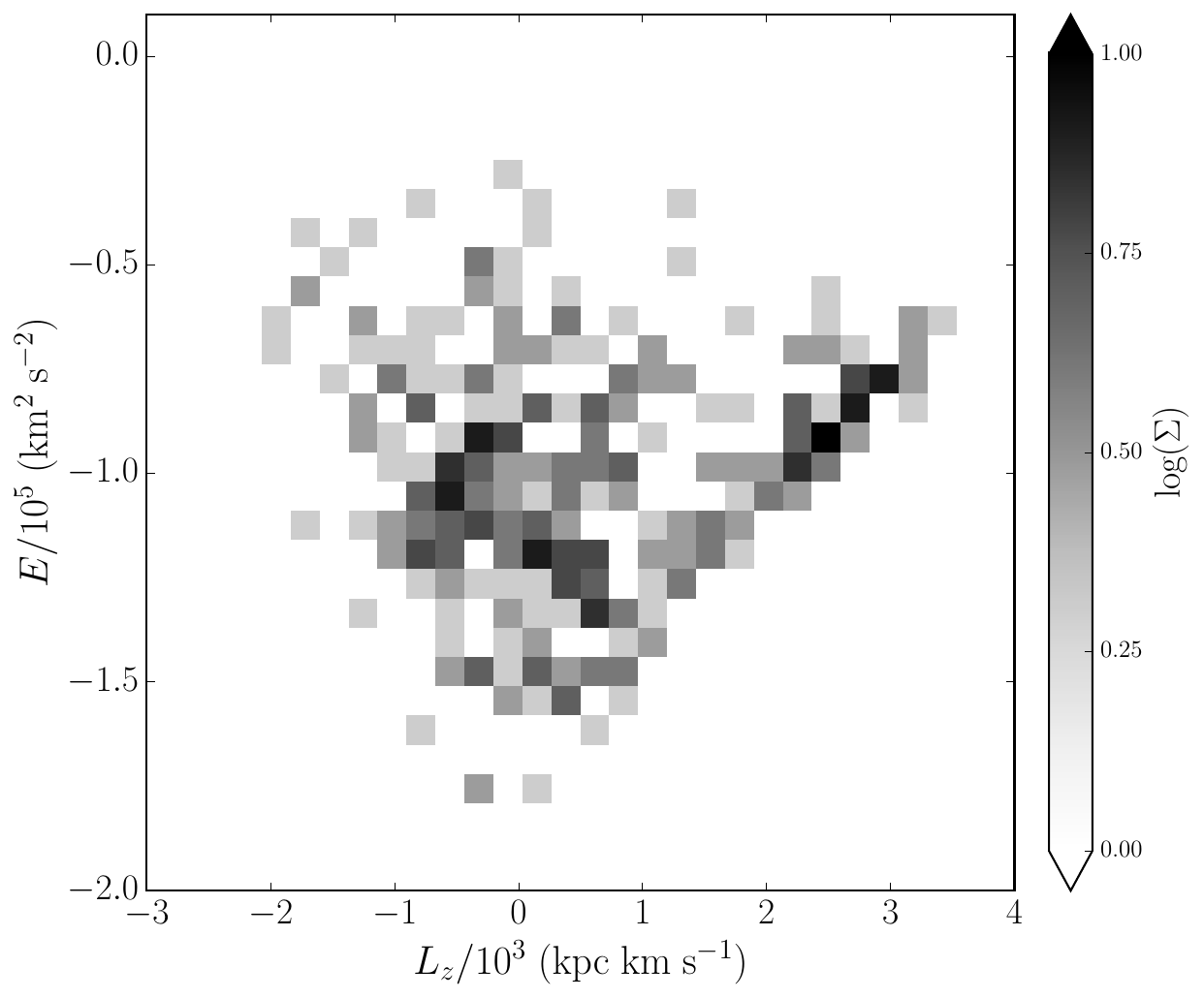}
        \vspace{0.5em} 
        \parbox[c]{\linewidth}{\centering \large R = 8 kpc}
        \label{fig:L4_sun}
    \end{minipage}
    
    \vspace{0em} 
    \caption{As the left panels in \figref{fig:e-lz_resonanece} but for the fiducial (Level 4) Auriga simulation of the same halo, these plots show the distribution of accreted stars in $E-L_z$ space in two spheres of radius 4 kpc at the centre of the halo and for a "solar neighbourhood" as defined in \figref{fig:sdens}. The stellar resolution of the fiducial Auriga simulation shown here is $\mathrm{5 \times 10^{4} \; M_{\sun}}$, which is 64 times less than Superstars. The overdensity caused by bar resonance, which is clearly visible in \figref{fig:e-lz_resonanece}, is completely absent at L4 resolution. The overdensity is absent as well in the in-situ population of L4 Auriga.}
    \label{fig:L4 comparison}
\end{figure}

Shown in \figref{fig:L4 comparison} is the distribution of accreted stars in $E-L_z$ space in the fiducial counterpart to the simulated halo we study in this paper. We select accreted stars at the central and "solar neighbourhood" regions. This figure serves as a direct comparison to the left-hand panels of \figref{fig:e-lz_resonanece}. We note that the ridge, which is very prominent in the Superstars halo, is not visible in the fiducial halo. Although there are differences stemming from different sets of random numbers in these two simulations, the complete absence of the ridge in the fiducial simulation is difficult to explain through these small differences. The increased stellar resolution in Superstars may bring out the ridge due to better sampling of phase space. Additionally, the large increase in stellar resolution leads to a more accurate potential in the inner regions of the halo, where stars dominate the potential. For a more detailed analysis of the effects of the Superstars method on the presence and nature of substructures we refer the reader to \citet{pakmor_2025}.

\section{The evolution of the ridge with time}

\begin{figure*}
    \centering
    \begin{subfigure}{\textwidth}
        \centering
        \includegraphics[width=.9\textwidth]{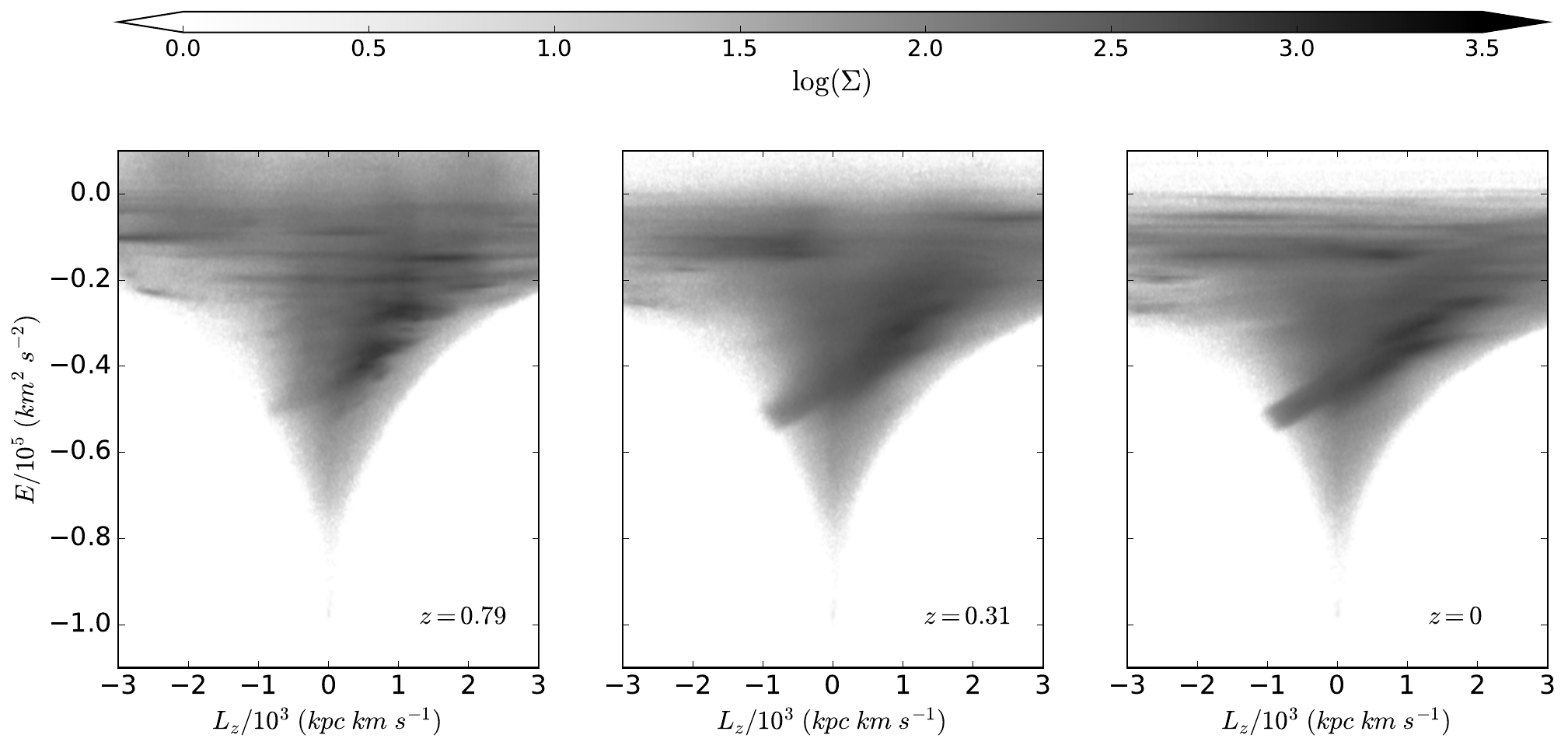}
    \end{subfigure}
    \caption{The total accreted population of the simulation, with no spatial cuts, in $E-L_z$ space at three different simulation snapshots as logarithmic density plots. The energy is normalised such that the minimum energy is $-1\times 10^5 \;\mathrm{km^2 \; s^{-2}}$ for all snapshots. The leftmost panel is at $z=0.79$, the centre panel at $z=0.31$, and the right panel at $z=0$. This corresponds to lookback times of 7 Gyrs, 3.6 Gyrs, and 0 Gyrs respectively. The ridge is visible in all plots, as all snapshots are after bar formation, however the ridge is getting stronger with time.}
    \label{fig:ridge_evolution}
\end{figure*}

An aspect of the ridge is that it becomes stronger with time. We show this by plotting the total accreted population of the simulation, with no spatial cut, in $E-L_z$ space at three different simulation snapshots at $z=0.79$, $z=0.31$, and $z=0$. This is shown in \figref{fig:ridge_evolution}. The ridge is visibly getting stronger with time, which is an indication that stars keep being trapped in resonance and are scattered into the ridge. The earliest of these snapshots is at a time around 1\,Gyr after the formation of the bar in this simulation, which occurred at a lookback time of 8\,Gyrs. Before bar formation the ridge is not visible.

\section{Alternative solar neighbourhood at R = 11.5\,kpc} \label{sec:appendix c}

\begin{figure*}
    \centering
    \begin{subfigure}{\textwidth}
        \centering
        \includegraphics[width=.9\textwidth]{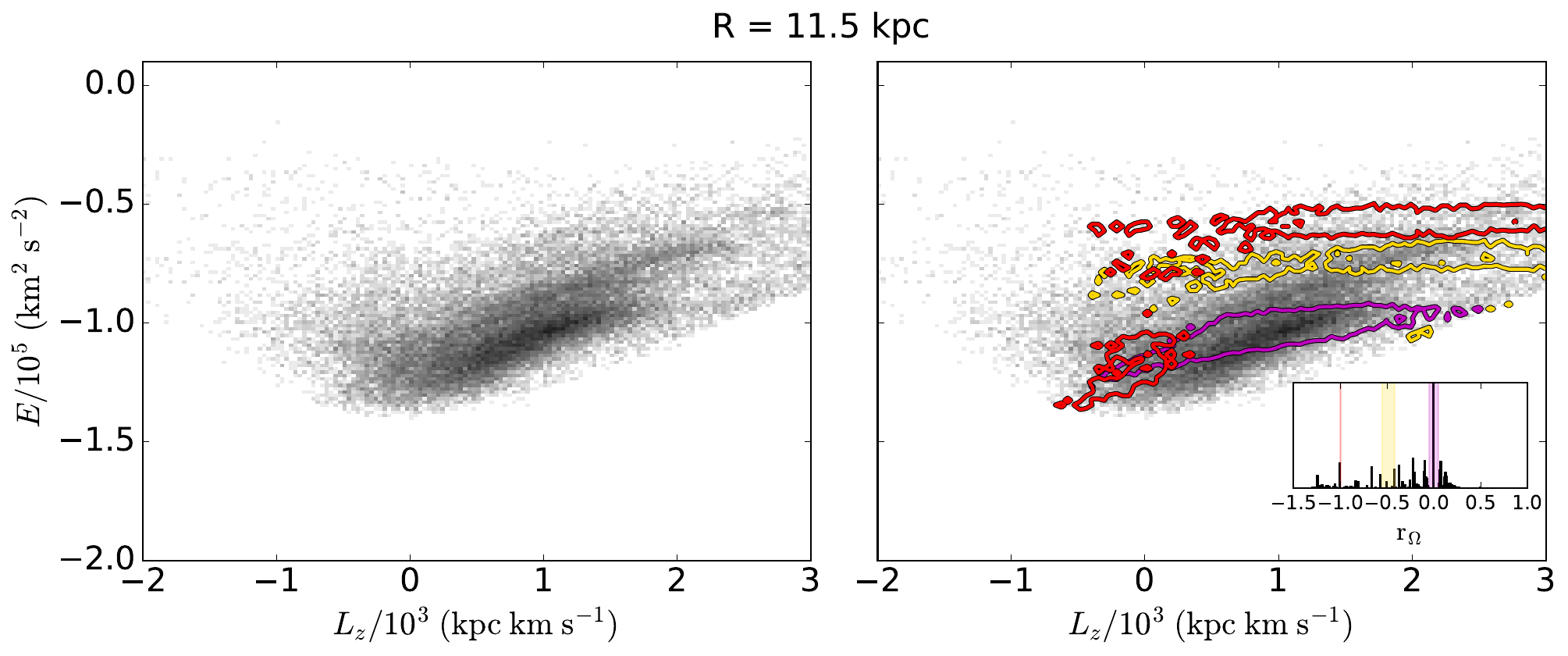}
    \end{subfigure}
    \caption{The ridge in $E-L_z$ space of the accreted population for a sphere placed at a radius of 11.5\,kpc rather than at 8\,kpc for our "solar neighbourhood". The left panel shows the distribution of accreted stars without contours, the right panel shows the same distribution with contours outlining the location of stars in resonance. The different colours denote three different resonances (CR: purple, 1:1: red, OLR: yellow). While the ridge is still visible, it is less pronounced for this region. Due to the larger corotation radius in the simulation compared to the Milky Way, this may be a closer resemblance of how this ridge may appear in observations.}
    \label{fig:other_sun}
\end{figure*}

Since the corotation radius of the simulation, at 10\,kpc, is larger than most estimates for the Milky Way (putting the corotation radius at $4.5-7\,$kpc \citet{bland_hawthorn_2016,sanders_2019,chiba_2021}), the spatial distribution of resonant stars is likely different in this simulation compared to the Milky Way. This will affect the visibility of the ridge feature. Thus, in \figref{fig:other_sun} we show the appearance of the ridge for a 4\,kpc sphere at a distance of 11.5\,kpc from the centre, 30 degrees behind the bar. This gives an impression of how apparent the ridge is in a region that may be more comparable to our solar neighbourhood in terms of the spatial distribution of resonant stars. We choose a radius of 11.5\,kpc by scaling up our simulation such that its corotation radius corresponds to a corotation radius of 7\,kpc in the Milky Way, i.e. $R_{\rm sun, \,new} = \frac{10}{7} \times 8\,\rm{kpc} \approx11.5\, \rm kpc$. We see that the ridge is still present as an overdensity; however, it is less distinct.

\end{document}